\documentclass[a4paper,11pt]{article}
\pdfoutput=1 

\usepackage{jcappub} 

\usepackage[T1]{fontenc} 
\usepackage{BOONDOX-cal}
\usepackage{soul}

\newcommand{\be}{\begin{equation}}
\newcommand{\ee}{\end{equation}}
\newcommand{\bea}{\begin{eqnarray}}
\newcommand{\eea}{\end{eqnarray}}

\newcommand{\df}{\dfrac}

\newcommand{\Fermi}{{\it Fermi}}
\newcommand{\Jf}{${J}$}

\newcommand{\sigmav}{\langle\sigma v\rangle}

\title{\boldmath Dark matter constraints from dwarf galaxies: a data-driven analysis}

\author{Francesca Calore,}
\author{Pasquale Dario Serpico,}
\author{and Bryan Zaldivar}

\affiliation{Univ. Grenoble Alpes, USMB, CNRS, LAPTh, F-74940 Annecy, France}
\emailAdd{calore@lapth.cnrs.fr}
\emailAdd{serpico@lapth.cnrs.fr}
\emailAdd{zaldivar@lapth.cnrs.fr}

\abstract{Dwarf galaxies represent a powerful probe of annihilating dark matter particle models, with gamma-ray data setting some of the best bounds available. A major issue in improving over existing constraints consists in the limited knowledge of the astrophysical background (mostly diffuse photons, but also unresolved sources). Perhaps more worrisome, several approaches in the literature suffer of the difficulty of assessing the systematic error due to background mis-modelling.  Here we propose a data-driven method to estimate the background at the dwarf position and its uncertainty, relying on an appropriate use of the whole-sky data, via an optimisation procedure of the interpolation weights. 
While this article is mostly methodologically oriented, we also report the bounds based on latest \Fermi-LAT data and updated information for \Jf-factors for both isolated and stacked dwarfs. Our results are very competitive with the \Fermi-LAT ones, while being derived with a more general and flexible method. We discuss the impact of profiling over the \Jf-factor as well as over the background probability distribution function, with the latter resulting for instance crucial in drawing conclusions of compatibility with DM interpretations of the so-called Galactic Centre Excess.}

\begin{document}

\begin{flushright}
LAPTH-010/18
\end{flushright}

\maketitle
\flushbottom

\section{Introduction}\label{introd}
Dark matter (DM) constitutes about 27\% of the matter-energy content of our Universe~\cite{Ade:2015xua} and 
discovering its nature is one of the biggest challenges of contemporary physics.
In the leading paradigm, DM is a new particle, beyond the framework of the standard model for particle physics. Several candidates in different ranges of mass and interaction strength exist and can account for the purely gravitational DM observational evidence.
One of the most promising class of candidates are weakly interacting massive particles
(WIMPs)~\cite{Bertone:2010zza}, which can naturally account for the observed DM cosmological relic density
with annihilation cross section $\langle\sigma v \rangle$ around the benchmark value $3\times 10^{-26}\,$cm$^3$/s~\cite{Steigman:2012nb}. Via their residual annihilations today, these 
DM particle candidates can produce stable particles such as photons and charged 
cosmic rays as a consequence of decay and hadronisation of the  
final states (quarks, gauge bosons, leptons, higgses).
The ensuing signal can, in principle, be detected by space- and ground-based telescopes. However, 
this possibility is challenged by the typically dominant astrophysical emissions, which one has to disentangle 
a WIMP signal from. High-energy gamma rays, from hundreds of MeV up to TeV energies, offer a channel of choice to perform such searches: on the one hand, peculiar spectral features in the (prompt) WIMP gamma-ray signal exist, which are unaffected by propagation/absorption effects at Galactic scales~\cite{Bringmann:2012ez}. On the other hand, the expected fluxes and the relative simplicity of detecting photons (as opposed to, say, neutrinos) allows one to collect significant statistics, potentially sufficient for a detection and a reliable discrimination from backgrounds. 

In the latest decade, the Large Area Telescope~\cite{2009ApJ...697.1071A}, 
aboard the \Fermi~satellite (hereafter \Fermi-LAT) has
collected gamma rays from hundreds MeV up to TeV energies from the whole sky, representing a unique 
instrument to probe high-energy non-thermal emission processes. 
Several targets in the sky might be and have been considered to perform DM (WIMP) searches~\cite{Conrad:2017pms}. 
One on the most promising targets are dwarf spheroidal galaxies (dSphs), satellites of the Milky Way located within a few hundreds kilo-parsecs (kpc) from the Galactic centre and whose mass is dominated by DM, as inferred from
the kinematics of their stars, see ref.~\cite{Mateo:1998wg} and references therein. 
With little intrinsic astrophysical background emission expected~\cite{Winter:2016wmy}, 
dSphs could be shining in gamma rays mostly in reason of their DM content, through annihilation (or decay) of WIMP particles.
So far, no significant excess of gamma rays has been found from the direction of known dSphs with the 
\Fermi-LAT, nor with TeV Cherenkov telescopes such H.E.S.S., MAGIC and VERITAS, and
upper limits on the strength of the annihilation cross section as a function of the WIMP mass have 
been placed~\cite{Ackermann:2015zua,Drlica-Wagner:2015xua,Fermi-LAT:2016uux,Abramowski:2014tra,Aleksic:2013xea,Ahnen:2016qkx,Archambault:2017wyh,Albert:2017vtb}, now challenging vanilla WIMP scenarios for masses up to $\sim 100\,$GeV. 
The search for DM in dSphs has received a significant boost in the past few years thanks to more than twenty newly discovered  objects -- both ``classical'' and ``ultra-faint'' dSphs~\cite{Bechtol:2015cbp,Koposov:2015cua,Laevens:2015kla,Laevens:2015una,Kim:2015ila,Kim:2015xoa,Drlica-Wagner:2015ufc} -- by wide-field optical imaging surveys such as SDSS, Pan-starrs and DES.
Follow-up spectroscopic observations, which are crucial for a precise determination of the dSphs DM content and its spatial distribution, are unfortunately difficult, in particular for the ultra-faint galaxies due to 
their very low surface brightness and small number of stars. This implies large uncertainties in the determination of their astrophysical $J$-factors, i.e.~the integrals along the line of sight of the DM density squared. 
Since the WIMP gamma-ray signal is directly proportional to this quantity, the constraints are affected correspondingly, see the discussion in ref.~\cite{Bonnivard:2015xpq}.

The pipeline of the standard analysis of DM searches in \Fermi-LAT dSphs is described in detail in ref.~\cite{Ackermann:2013yva}.
The main idea is to look for gamma-ray point-like source emission from the direction of each dSph, and then, to stack the (null-)results from the whole sample of objects in a joint likelihood, eventually accounting for uncertainties on their DM content via profiling over the dSphs \Jf-factor. 
The background in a $15^\circ \times 15^\circ$
region of interest (ROI) centred at the dSph position is determined through a template fit of the \Fermi-LAT diffuse
and isotropic emission models, known \Fermi-LAT sources in the ROI and, in addition, the new dSph point-like source.
Although powerful for setting constraints, this method suffers from some limitations. In particular, 
the background is fitted in each ROI independently, hence there is no guarantee that it is consistently determined from one region to another. Also, the capability of the model in reproducing the background in the ROI depends on the inclusion of a satisfactory number of modelled astrophysical components. For instance, the presence of a new spatially-dependent unresolved population of objects may provide unequal performances in different regions of the sky, if it is unknown and unaccounted for. 
It is also difficult to assess the systematic error associated to the theory biases implicitly hidden in the above procedure. The \Fermi-LAT team estimates this by studying the impact of different background modelling on the final bounds, but it is certainly not a statistically sound procedure. Alternatives have been proposed, usually oriented towards a more data-driven approach for the background estimate. For instance, in the methods proposed in~\cite{GeringerSameth:2011iw} (also adopted in the recent~\cite{Boddy:2018qur}) and~\cite{Mazziotta:2012ux} the counts in a background region around the source are used to construct an empirical PDF of the background at the source, essentially by rescaling for the solid angle ratio. Additionally,~\cite{Mazziotta:2012ux} proposed a generalisation to multiple energy bins and a Bayesian approach, as well as a treatment of \Jf-factor uncertainties with top-hat, flat-priors,  vs.~a cumulative and frequentist approach in~\cite{GeringerSameth:2011iw}, with a simplified treatment of \Jf-factor uncertainties. While they do not rely on astrophysical modelling, these techniques also suffer from similar drawbacks: for instance, the background region around each dSph is somewhat chosen arbitrarily. Also, all the points within it are equally important, while one may physically expect that directions closer to the dSph should be more determinant for the background estimate at the dSph. Additionally, it is hard to evaluate how performing the method is. These approaches have also little to say on strategies for improvements.

In an attempt to cope with these limitations, we asked ourselves if a data-driven approach could provide a complementary strategy, in turn allowing one to investigate the robustness of the constraints.  Our aim is to develop a general method to model the background at all dSphs positions simultaneously and consistently using the whole sky information, based on machine learning methods rather than explicit astrophysical modelling.
As shown in ref.~\cite{Charles:2016pgz}, DM searches in dSphs are background-limited for low DM masses 
($\lesssim$ 100 GeV). We therefore expect our method to provide a better assessment of the actual limits, especially at low DM masses. Our main goal is methodological, and we expect the method to be useful for background estimations also for generic \Fermi-LAT targeted searches. 
With future developments in mind, in the following we will also highlight some simplifying hypotheses that we made which are susceptible to further refinements. 

The paper is organised as follows:  section~\ref{sec:DMsignal} describes how we model the DM signal, 
following traditional and well-known prescriptions. 
In section~\ref{sec:KS} we present the data set used and the procedure adopted to generate control regions.
In section~\ref{sec:likelihood} we proceed in setting upper limits on a possible WIMP signal in dSphs 
developing a likelihood analysis, based on the background determination previously described in section~\ref{sec:Method}.
We present our results in section~\ref{sec:results} and discuss our conclusions and some perspectives in section~\ref{sec:conclusions}.

\section{Dwarf spheroidal galaxies: DM signal and catalogue}
\label{sec:DMsignal}
We model the DM signal from each dSph as conventionally done in the literature.
The gamma-ray flux (per unit energy) from self-annihilating Majorana DM particle 
candidates from the direction of each dSph writes as:
\begin{equation}
\frac{{\rm d}\Phi}{{\rm d}E} = \frac{ \sigmav }{8\pi m_{\rm DM}^2} \; \frac{dN_\alpha}{dE} \; {J} \, , 
\label{eq:phflux}
\end{equation}
where $\sigmav$ is the velocity-averaged annihilation cross section, $m_{\rm DM}$ is the DM particle mass, 
$dN_\alpha/dE$ is the DM annihilation spectrum, providing the number of photons resulting from the annihilation
of a DM pair in a pair of final state particles of species $\alpha$. As benchmark, we here assume 100\% branching fraction
into $b$-quarks, and the DM spectrum is derived by using the PPPC4DMID~\cite{Cirelli:2010xx}.

Finally, the so-called $J$-factor corresponds to the integral along the line of sight of the DM density squared.
For dSphs with spectroscopic information, the \Jf-factor can be derived  from an analysis of the kinematics of stars
in the dSphs. Measured $J$-factors (and their uncertainties) are available for classical dSphs and for some ultra-faint objects.
As a reference, we use the values quoted in table~1 of ref.~\cite{Fermi-LAT:2016uux}, which correspond to $J$-factors within a circular 
region of 0.5$^\circ$ radius.  Since the angular size ($\theta_s \sim r_s/d$) of the dSph is contained within this range, it encompasses almost all the signal emitted. At the same time, this size roughly matches the ``median'' energy-dependent point-spread function of the \Fermi-LAT, hence minimizing the effects of leakage of the signal outside the search window, and greatly simplifying the analysis. This explains why this size choice has become a standard lore in the literature, and why we stick to it in order to simplify both the analysis and the comparison with previous results.
To obtain the photon counts at each dSph position, we integrate the differential gamma-ray flux in eq.~(\ref{eq:phflux})
in the desired energy bin, $e$, convolving with the exposure map, $\mathcal{E}_d(E)$, at the corresponding dSph position, labelled by $d$.
We can write the expected DM counts from each dSph in the energy bin $e$ as:
\begin{equation}
\lambda_{d, e}^{\rm DM} = \int_e \frac{{\rm d}\Phi}{{\rm d}E} \; \mathcal{E}_d(E) {\rm d}E\,  \equiv J_d \; \sigmav \; f_{d,e}(m_{\rm DM}) \, ,
\label{eq:phcounts}
\end{equation}
where the function $f_{d,e}(m_{\rm DM})$ encodes the spectral information of the signal and the convolution with the exposure map.

In the present analysis we use a sample of 25 dSphs, all of which have \Jf-factors estimated from spectroscopic measurements. In addition to the 19 kinematically confirmed galaxies with \Jf-factors quoted in table~1 of ref.~\cite{Fermi-LAT:2016uux}~\footnote{These \Jf-factors were derived by stellar kinematics in ref.~\cite{Geringer-Sameth:2014yza}.}, we consider Horologium I, Hydra II, Pisces II, Willman I and Grus I using the \Jf-factor estimates in ref.~\cite{Evans:2016xwx}, and Tucana II with the measured \Jf-factor from ref.~\cite{Walker:2016mcs}. 
The robust identification of satellites as well as the measurement of the \Jf-factor from stellar kinematics is very delicate, in particular for ultra-faint galaxies where contamination effects and non-equilibrium dynamics can be difficult to account for. A thorough discussion of the \Jf-factor determination and its uncertainties is beyond the scope of the present paper, pointing the interested  reader to recent relevant literature in this respect, see e.g.~\cite{Bonnivard:2015xpq,Geringer-Sameth:2014yza,Evans:2016xwx}. 

\section{Gamma-ray data selection}
\label{sec:KS}

Our main assumption is that the (putative) DM signal might come from circular regions, centred at the positions of the 25 dSphs with known \Jf-factor and with a radius of 0.5$^{\circ}$, and that the rest of the sky is signal-free or, better, background-dominated.  A few comments on these approximations are in order: 
first, the background sample is expected to contain not only photons from standard astrophysical sources but also from DM annihilation in the halo (the same is obviously true for the \Fermi-LAT procedure). The fact that the diffuse Galactic emission agrees within few tens of percent with astrophysical model expectations~\cite{2012ApJ...750....3A} suggests that, globally, the DM halo signal constitutes at most ${\cal{O}}$(10\%) of the astrophysical background. The contribution of the DM halo to the DM flux coming from the much denser sub-halos hosting the dwarfs galaxies should also be sub-leading. 
Additionally, this is not a concern for our method, which implicitly accounts for this component in the overall background expectation (both background regions and dwarfs galaxies are superposed to the background DM halo).
Nonetheless, one might envisage to generalise our approach to account for an additional, diffuse DM signal, whose morphology is roughly known, and with an (almost) identical spectrum contributing to both samples. This simultaneous halo-dSph DM analysis might further tighten the sensitivity of gamma-ray data to DM, but is left for future work.

Secondly, among the background regions, some will lie along the direction of additional dSphs, yet to be discovered.  Thus, one expects a partial contamination of the void sample by actual dSph signal regions. 
To mitigate such an effect to the best of our actual knowledge, we actually mask the direction towards the complete sample of 44 dSphs (confirmed and candidates) of ref.~\cite{Fermi-LAT:2016uux} --despite using only the 25 listed in sec.~\ref{sec:DMsignal} as our candidate signal-- in constructing our background region (see below).
However, even ${\cal{O}}$(100) more dSphs to be discovered (already optimistic compared to current expectations~\cite{Charles:2016pgz}) constitute but a small fraction of the void sample considered here, amounting to tens of thousands of line-of-sight pixels. Therefore we expect a negligible bias in our procedure. 

We use \Fermi-LAT public data collected over almost seven years (from August 4, 2008 to June 13, 2015). In particular, we rely upon the SOURCE \Fermi-LAT P8R2 class data, including both front- and back-converted events in the energy range from 500 MeV to 500 GeV.  Standard cuts to events as suggested by the \Fermi~Science Support Center
(``(DATA\_QUAL$>$0) \&\& (LAT\_CONFIG==1)'' and zenith angles $> 90^\circ$) are applied. 
We consider all-sky data, binned in energy in 24 log-spaced bins and 
spatially binned in Cartesian coordinates with pixel size of 0.1$^\circ$. 

\begin{figure}[h]
  \centering
  \includegraphics[scale=0.65]{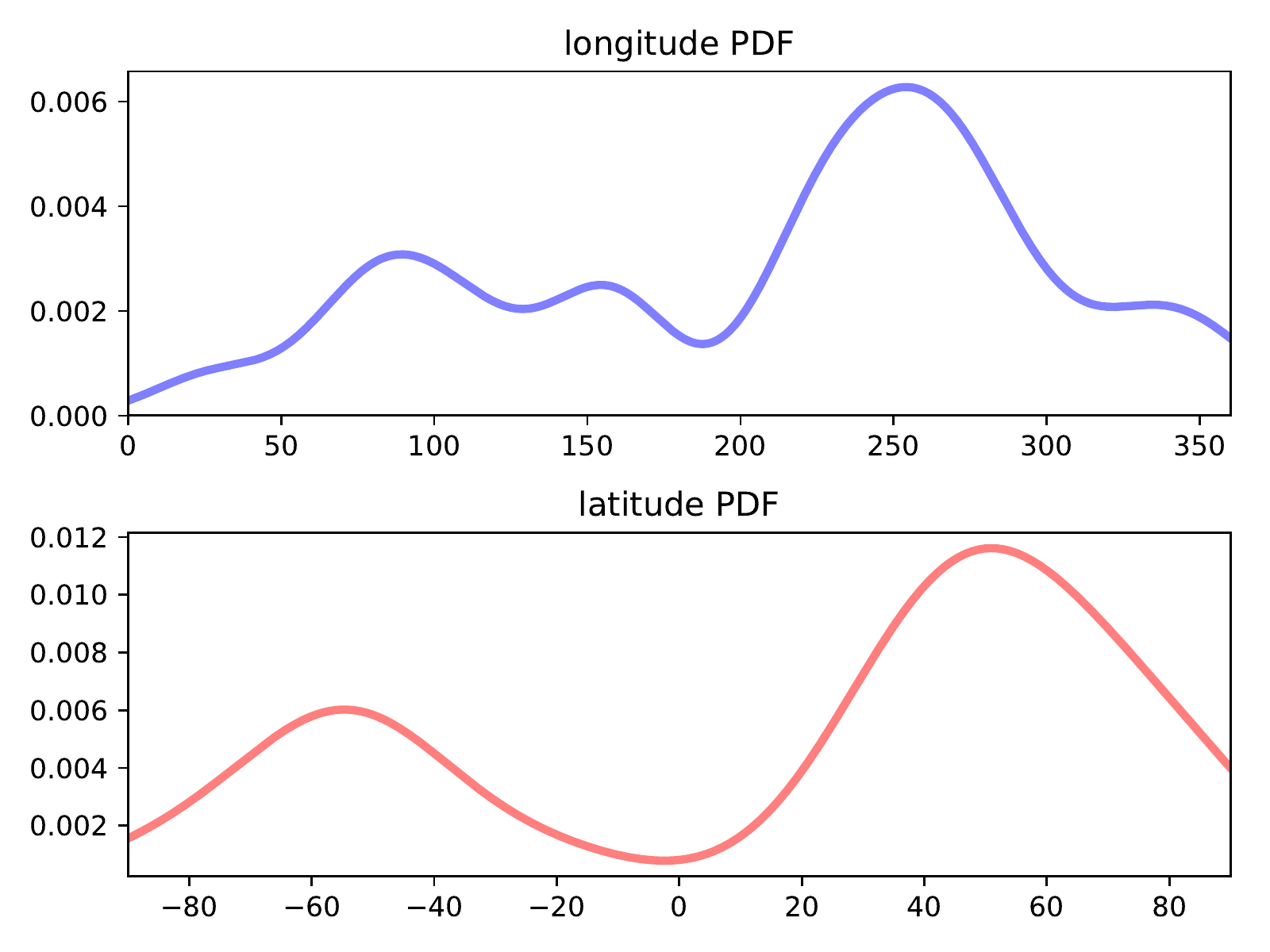} 
  \caption{\it Probability distribution functions of Galactic longitude and latitude (in degrees) of the 25 dSphs selected in the present analysis.
  }
\label{fig:PDFs}
\end{figure}

In the whole sky, we randomly generate the maximum possible number of background-dominated (i.e.~signal-free) circular regions (hereafter called ``void regions'') of the same size as the signal (dSphs) regions 
and with the same spatial distribution (in Galactic longitude $|\mathcal{l}|$ and latitude $|\mathcal{b}|$) of our 25 dSphs. This distribution is thought to
be dominated by observational limitations, hence the extraction of a similarly distributed background is required to minimise possible biases. 
To extract separately the longitude and latitude distribution functions of the 25 dSphs\footnote{We here introduce the simplifying assumption that the two variables are uncorrelated.}, we apply the Kernel Density Estimation (KDE) technique, with the respective bandwidths fixed to be the values best fitting the data, according to the Grid Search method in the {\tt scikit-learn} {\scshape Python}'s package. The resulting optimum bandwidths in degrees are 20.0 (15.6)  for the longitude (latitude) distribution, which gives rise to the
probability density function (PDF) of the dSphs positions shown in figure~\ref{fig:PDFs}.

 \begin{figure}[h]
  \centering
  \includegraphics[scale=0.6]{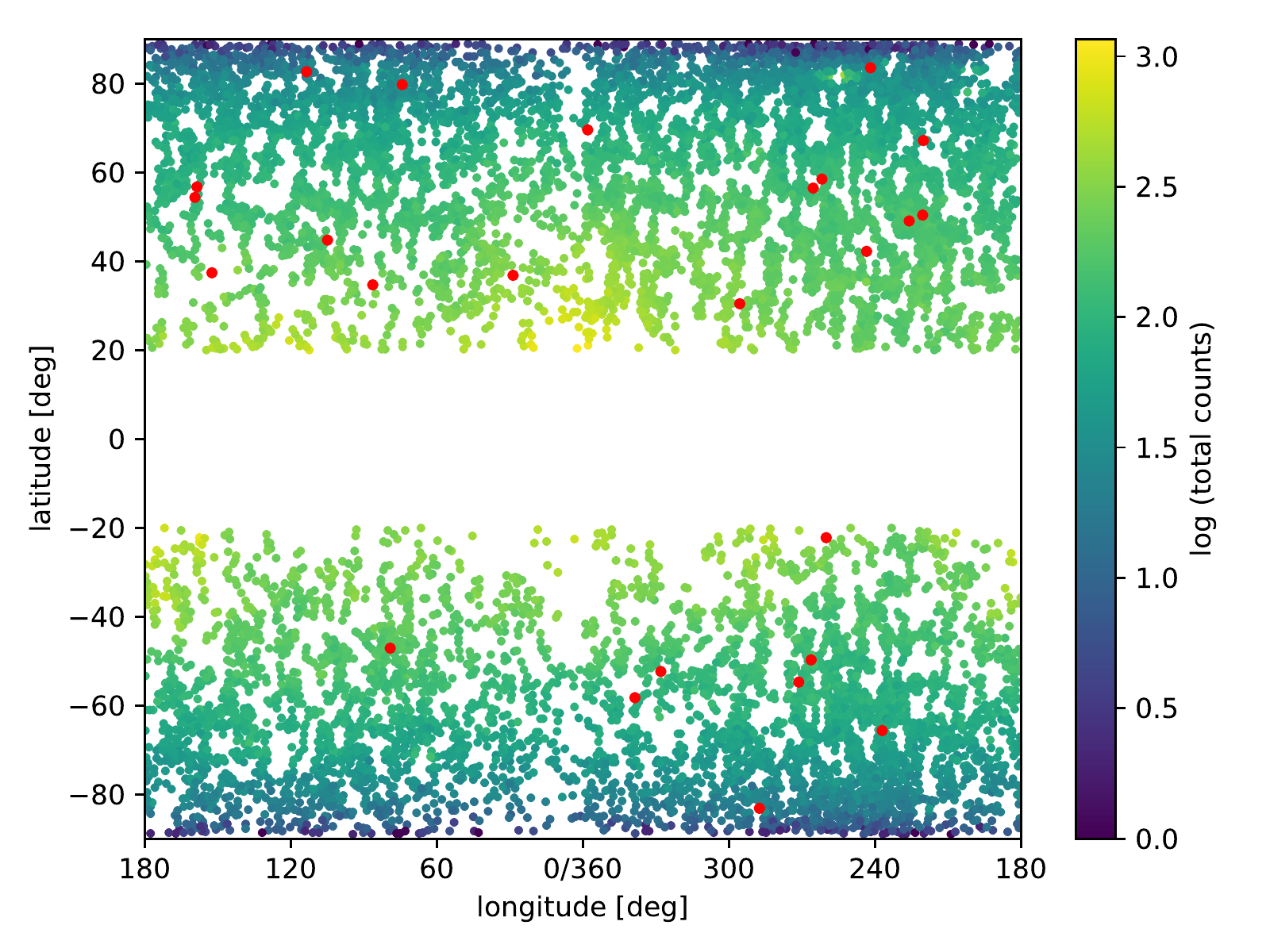}
  \caption{\it Galactic coordinate map of the generated background regions (see text for the details of the procedure). The colour represents the $\log_{10}$ of the number of counts. The red dots are the 25 dwarfs positions to be used in our analysis (sizes are modified for the sake of visibility).}
\label{fig:map}
\end{figure} 

From such a first set of void regions, we remove those lying along the disk ($|\mathcal{b}|\leq 20^\circ$), those coincident with the third \Fermi-LAT's catalog of point-like as well as extended sources~\cite{Acero:2015hja}, and, finally, those coincident with all the 44 dSphs (confirmed and candidates)~\cite{Fermi-LAT:2016uux}. After this procedure, around $N\sim 9400$ void regions remain which constitute our  background-only sample, see figure~\ref{fig:map}.

\bigskip

\section{Background prediction at the dSphs positions}
\label{sec:Method}
A naive application of the most frequently used machine-learning approaches to our problem would consist in assuming a mapping $b(\vec x,\vec \beta)$ between the directions on the sky $\vec x = (\mathcal{l},\mathcal{b})$ and the photon counts due to astrophysical background $b$ measured by \Fermi-LAT, depending on a number of parameters $\vec \beta$.
In turn, the parameters  are fixed through some adaptive scheme on a sub-sample (so-called {\it training} sample) 
 of the data $b_i$ corresponding to positions $\vec x_i$. The extension to multiple energy-bins is  straightforward, with the counts becoming a vector $\vec b$.

However, such an estimator assumes an underlying deterministic map, whereas we know that $b$ is a stochastic variable (due to Poisson fluctuations, irregularly sampled time dependent signals, etc.). For this reason, from a physics point of view, it is more meaningful to build an estimator of parameters of {\it the distribution} of $b$, instead of estimating $b$ itself.  This strategy can be implemented with parametric or non-parametric methods. For completeness, in the subsec.~\ref{parmet}, we briefly illustrate the former, although we opt for the latter strategy, detailed in subsec.~\ref{nonparmet}. The main advantage of a non-parametric method is that it removes some arbitrarity in the choice of the PDF of the background, since under quite general conditions one is guaranteed (at least asymptotically) to converge to an unbiased estimator for the true underlying PDF. The free parameters entering the form of the Kernel of this estimator are then fixed according to a data-driven optimization, which in principle involves all points in our sample. Subsec.~\ref{sec:Ebin} finally explains some technicalities on the extension of the method to an energy-dependent context, together with some approximations which have a minor impact, but greatly simplify an otherwise too heavy numerical problem.

\subsection{Parametric methods}\label{parmet}
In a {\it parametric} likelihood approach, one would for instance assume that $b$ obeys a PDF ${\cal F}$ (e.g. Gaussian) via some PDF parameters (e.g. mean $\sf b$ and variance $\sigma^2$), which are in general functions of the directions $\vec x$, controlled by some parameters $\vec \beta$ to be optimised on the data.  For instance one may assume that data are drawn from identical-variance distributions (i.e. $\sigma(\vec x)=\sigma$) but position-varying mean, and adopt a basis function method \cite{ESL} to write:
\be
{\sf b}(\vec x)=\sum_i^p \beta_i h_i(\vec x)~,
\label{MLex}
\ee
where $p$ is the number of optimisation parameters as well as of the ``response'' (in general non-linear) functions $h_i(\vec x)$ to be appropriately chosen.
Examples of bases include polynomials of degree $i$ or Gaussians centred around different points. 

The learning phase consists in finding, for a given value of $p$, the value of the parameters $\beta_i$ by maximising the product likelihood $\Pi_\alpha {\cal F}(b_\alpha,\vec x_\alpha)$ evaluated on a chosen {\it training} sample  of size $N$, with $\alpha$ an index within this sample. In this specific case it can be proven~\cite{ESL} that this is achieved for $\sf b\to {\sf \hat b}$ corresponding to $\vec\beta=\hat{\vec \beta}$ given by 
$\hat{\vec \beta} = (H^TH)^{-1}H^T \vec b$, where $H_{\alpha i}\equiv h_{i}(\vec x_\alpha)$ is a $N\times p$ matrix  and $\vec b$ ~is the $N$-component vector of counts. In a second step, one substitutes the  ``optimal'' values $\hat{\vec \beta}$ (they can be in general {\it estimated} with clever numerical techniques, rather than analytically as above) on the same likelihood as before, but now evaluated on an independent (called ``test'') sample, whose result is stored. One performs the above procedure for a series of values of $p$. Finally, the optimal model (in this case, the optimal value of $p$) is chosen to be the one giving the maximum likelihood over the test sample.    Most of the machine-learning methods follow this approach, including, for example, other nonlinear parametric models such as neural networks. 
   
\subsection{Non parametric methods}\label{nonparmet}
The above strategy requires an assumption for the PDF of the background count variable. Rather, we will follow a kernel (also called ``non-parametric'') estimation for the PDF of the ``output'' $y$,    
\be 
{\hat {\cal F}}(\vec x, y) = \frac{1}{N}\sum_{i=1}^{N} {\cal f}(\vec x-\vec x_i, y- y_i) \, , 
\label{kernelPDF}
\ee  
where we adopt a factorised form depending upon two ``bandwidth''  parameters $\sigma,\,\varsigma$ 
\be 
 {\cal f}(\vec x-\vec x_i, y- y_i)=K_\sigma(\vec x-\vec x_i)g_\varsigma( y- y_i)\,.\label{kernelFact}
\ee  
Under quite weak hypotheses of continuity and smoothness (see the original work by Parzen~\cite{parzen1962} for details),
the choice of eq.~(\ref{kernelPDF}) is guaranteed to provide an unbiased estimator for the true underlying PDF, ${\cal F}$, in the limit of large-$N$, independently of the specific choice of the kernel.
For later analytical convenience, we choose a log-normal distribution for the counts' kernel: 
\be 
g_\varsigma(b,b_i)= \frac{1}{\sqrt{2\pi}\,\varsigma
\,b}\exp\left[-\df{(\ln b- \ln b_i)^2}{2\varsigma^2}\right]\,,
\label{kernelY}
\ee  
which is thus centred around $\ln b_i$. This choice of functional dependence is physically motivated by our goal of obtaining an estimator whose {\it relative} error (as opposed to absolute error) on the counts is homogeneous across the sky.

In the following, we adopt a Gaussian form for  the function $K_\sigma$, 
\be 
K_\sigma(\vec x,\,\vec x_i)=\df{1}{2\pi\,\sigma^2}
\exp\left[-\df{(\vec x - \vec x_i)^T (\vec x - \vec x_i)}{2\sigma^2}\right]\,. 
\label{Kest}
\ee
While being a popular choice, it is certainly not unique, see \cite{bishop} for the properties of kernel constructions.  

The true conditional mean $  \ln {\sf b}(\vec x)$ writes in general and for arbitrary normalization of ${\hat {\cal F}}$ as
\be 
\ln {\sf b}(\vec x)\equiv 
 \df{\int_{-\infty}^{\infty} \ln b \, {\cal F}(\vec x, b) {\rm d}b}{\int_{-\infty}^{\infty} {\cal F}(\vec x, b) {\rm d}b}\,.
\ee 
Adopting ${\hat {\cal F}}(\vec x, b)$ as from eqs.~(\ref{kernelPDF})-(\ref{kernelFact})
as PDF yields the estimate $\widehat{\ln {\sf b}}(\vec x)$:
\be 
\widehat{ \ln {\sf b} }= \df{\sum_{i=1}^n K_i\ln b_i}{\sum_{i=1}^n   K_i}\,,
\label{estim}
\ee 
where we adopted the short-hand notation $K_\sigma(\vec x,\,\vec x_i)=K_i$. It will be useful for later convenience to define the variance of $\ln b$, which depends instead on both $\sigma$ and $\varsigma$ and writes:
\be 
{\rm Var}(\ln b)_{\sigma,\,\varsigma} \equiv
{\rm E}[ ( \ln b - {\rm E}[ \ln b]) ^2]
=\widehat{ \ln {\sf b}^2 }- (\widehat{ \ln {\sf b}})^2\, \, ,
\label{Var}
\ee
where ${\rm E}[\ldots]$ indicates the expectation value of its argument in brackets. 
A direct evaluation of the above expression using the adopted PDF form, eq.~(\ref{kernelPDF}), yields:
\be 
\widehat{ {\rm Var}}(\ln b)_{\sigma,\,\varsigma} =
\varsigma^2+\left[\frac{\sum_{i=1}^{n} K_i (\ln b_i)^2}{\sum_{i=1}^{n} K_i}-\left(\frac{\sum_{i=1}^{n} K_i \ln b_i}{\sum_{i=1}^{n} K_i}\right)^2\right]~.
\label{VarEstim}
\ee

Eq.~(\ref{estim}) is the defining property of the {\it Nadaraya-Watson} kernel model (see~\cite{ESL,bishop}). This is one of the simplest examples of ``memory-based'' methods that imply storing the whole dataset in order to make predictions for future data points. In other words, the whole dataset is considered as a ``training'' sample, in contrast with the training/test interplay described in the previous section. Consequently, there is a conceptual difference between eq.~(\ref{estim}) and an estimate based on eq.~(\ref{MLex}): in eq.~(\ref{estim}) the output enters explicitly the estimator, while eq.~(\ref{MLex}) once optimised/trained, does not require storage of the output values of the training sample. 

The use of eq.~(\ref{estim}) in regression problems for machine-learning applications was, up to our knowledge, first proposed by Specht in~\cite{GRNN}, where two important simplifications were considered: (a) the kernels $K_\sigma$ and $g_\varsigma$ are chosen to be identical, and most importantly (b) the bandwidth parameters are chosen to be equal, $\sigma=\varsigma$~\footnote{Also, in \cite{GRNN}, a Gaussian instead of a log-normal distribution is used. Note however that the result in eq.~(\ref{estim}) is the same, since the estimator $\widehat{\ln{\sf b}}$ does not depend on the choice of the function $g$.}. For our purposes, those simplifications are not convenient, since the two quantities the PDF in eq.~(\ref{kernelPDF}) depends upon are dissimilar. Additionally, we adopt a search criterion for the optimal parameters $\sigma$ and $\varsigma$ which is more generic than the one proposed in \cite{GRNN}, as we describe below.  
\newline\newline
In order to estimate the optimal values of $\{\sigma,\,\varsigma\}$ we follow the Maximum Likelihood Estimate (MLE) method. According to eqs.~(\ref{kernelPDF})-(\ref{Kest}), and assuming all the points in the sample as statistically independent, we build the (logarithm of the) total likelihood as:
\be 
\ln \hat{\cal F}_{\rm tot}(\sigma,\varsigma) = \sum_{i=1}^N \ln \left[\frac{1}{N-1}\sum_{j\neq i}^{N-1}K_\sigma(\vec x_i,\vec x_j)g_\varsigma(b_i,b_j) \right]~.
\label{totPDF}
\ee
In other words, we take the  total likelihood to be the product of the  likelihood at each point $(\vec x_i,b_i)$, and the latter is the PDF constructed from the remaining $N-1$ data points, evaluated at $(\vec x_i,b_i)$. We scan $\ln \hat{\cal F}_{\rm tot}(\sigma,\varsigma)$ over $\{\sigma,\,\varsigma\}$ and extract:
\be 
\{\sigma^*,\varsigma^*\} = 
{\rm argmax}~\ln \hat{\cal F}_{\rm tot}(\sigma,\varsigma) \approx \{1.58,\,0.16\}~.
\label{sigvarsig}
\ee 
 \begin{figure}[h]
  \centering
  \includegraphics[scale=0.6]{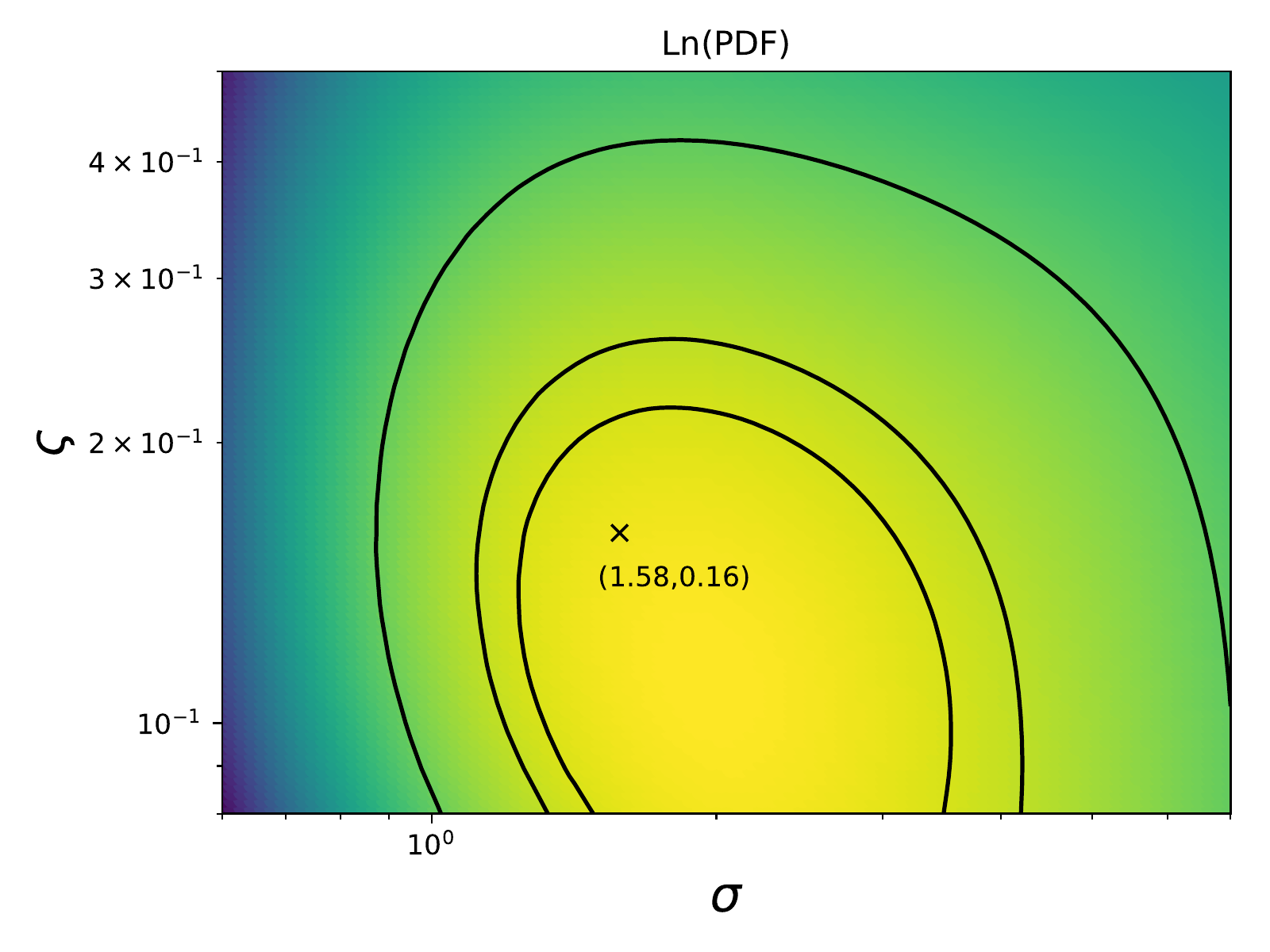}
  \caption{\it Scan of the total PDF (eq.~\ref{totPDF}) as a function of its parameters $\sigma$ and $\varsigma$. The {\tt X} symbol and the text specify the position of the maximum. For illustration purposes only, we also report curves representing values 1.005, 1.01 and 1.03  times (from inner to outer isocontours, respectively) the maximum of $\ln \hat{\cal F}$ (which is negative).}
\label{fig:PDFtot}
\end{figure} 
The result is shown in fig.~\ref{fig:PDFtot}, together with the position of the maximum. We note that the optimum $\sigma$ value, $\sigma^*\approx 1.58^\circ$, corresponds to small angular distances. On the one hand, this justifies a posteriori the fact that we used a flat-sky ``metric'' in order to compute distances between points, since those for which this is a bad approximation do not really contribute to the evaluated quantities. On the other hand, it means that only the close neighbours of each point $i$ really contribute to the PDF at  this location, something which is a consequence of having noisy data. Note also that although we have extracted global parameters, a priori the values of $\{\sigma,\varsigma\}$ that maximise the PDF at each point $i$ may be different than $\{\sigma^*,\varsigma^*\}$. For example, in the left panel of fig.~\ref{fig:PDFi} we plot the PDF at a particular location among the measured ones, $\vec x = (322.4^\circ, 63.7^\circ)$, with corresponding measured count rate $b=29$, as a function of $\sigma$ for three values of $\varsigma$. 
It is visible that the maximum is shifted by an ${\cal O}(1)$ factor with respect to $\sigma^*$ in (\ref{sigvarsig}). 
Finally, in order to illustrate the effect of $\varsigma$ on the width of the PDF,  in the right panel of fig.~\ref{fig:PDFi} we plot the function vs the counts, evaluated at the same point as in the left panel while fixing $\sigma=\sigma^*$, for the same three values of $\varsigma$. In the limit of small values for $\varsigma$, the PDF assumes the aspect of a comb-like function, being the superposition of narrow kernels each associated to the contribution of one neighbouring point. As a result of (multiple) local minima, the evaluation of the function at intermediate count values with respect to the values attained in the neighboring points may be low: this is a situation of comparatively small variance with potentially large bias-to-variance ratio. In the opposite limit of large $\varsigma$, the PDF is so broad that the function is certainly smooth and thus the bias-to-variance ratio is minimal, but the likelihood near the maximum has clearly worsened (situation of a too large variance). The optimum we find, $\varsigma^*=0.16$, is a manifestation of the {\it bias-variance} tradeoff, with a balanced choice being recommended in the specialised literature as desirable feature of an appropriate optimisation criterion (see for instance~\cite{ESL}).

 \begin{figure}[h]
  \centering
  \includegraphics[scale=0.44]{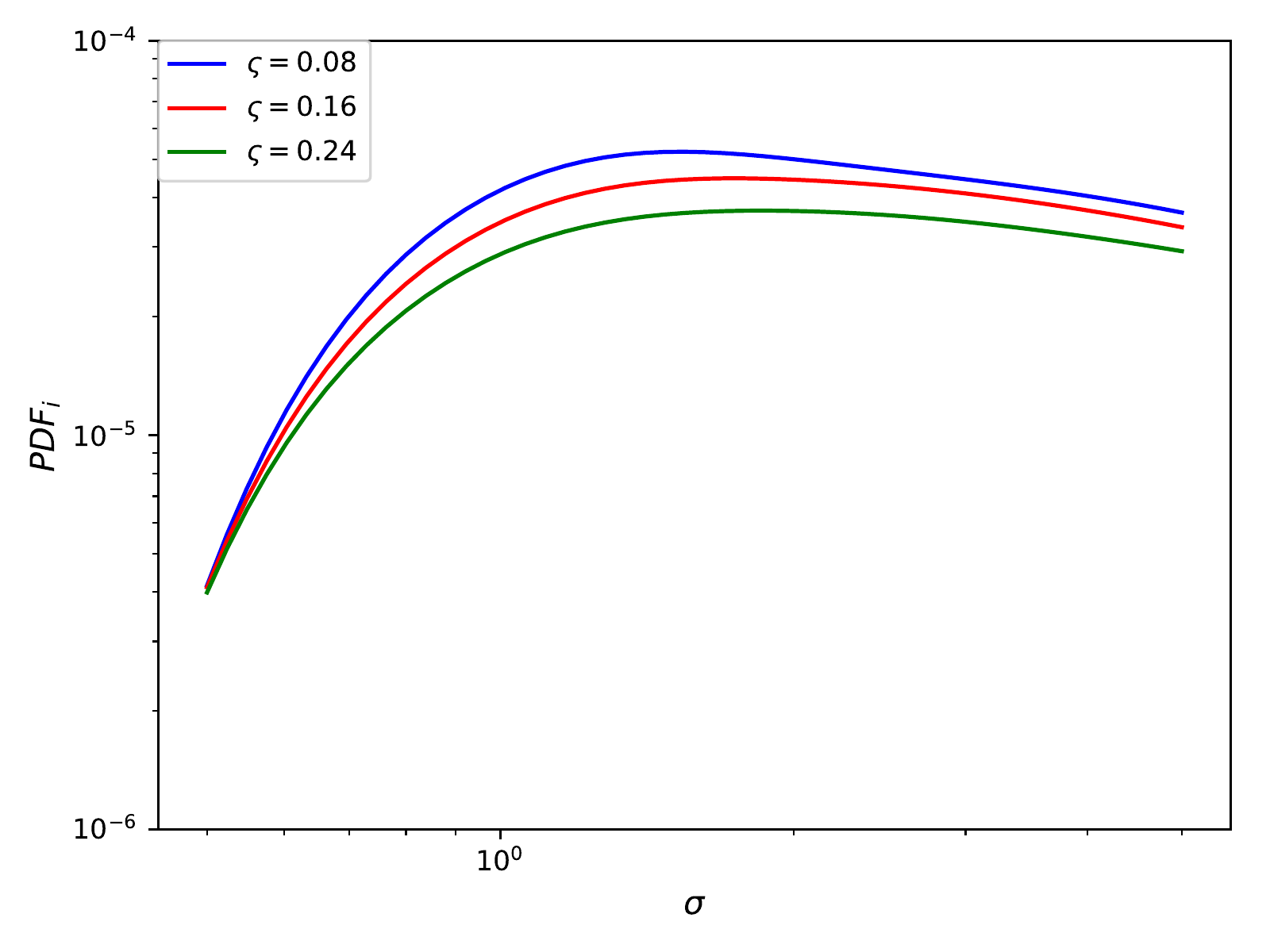}
  \includegraphics[scale=0.44]{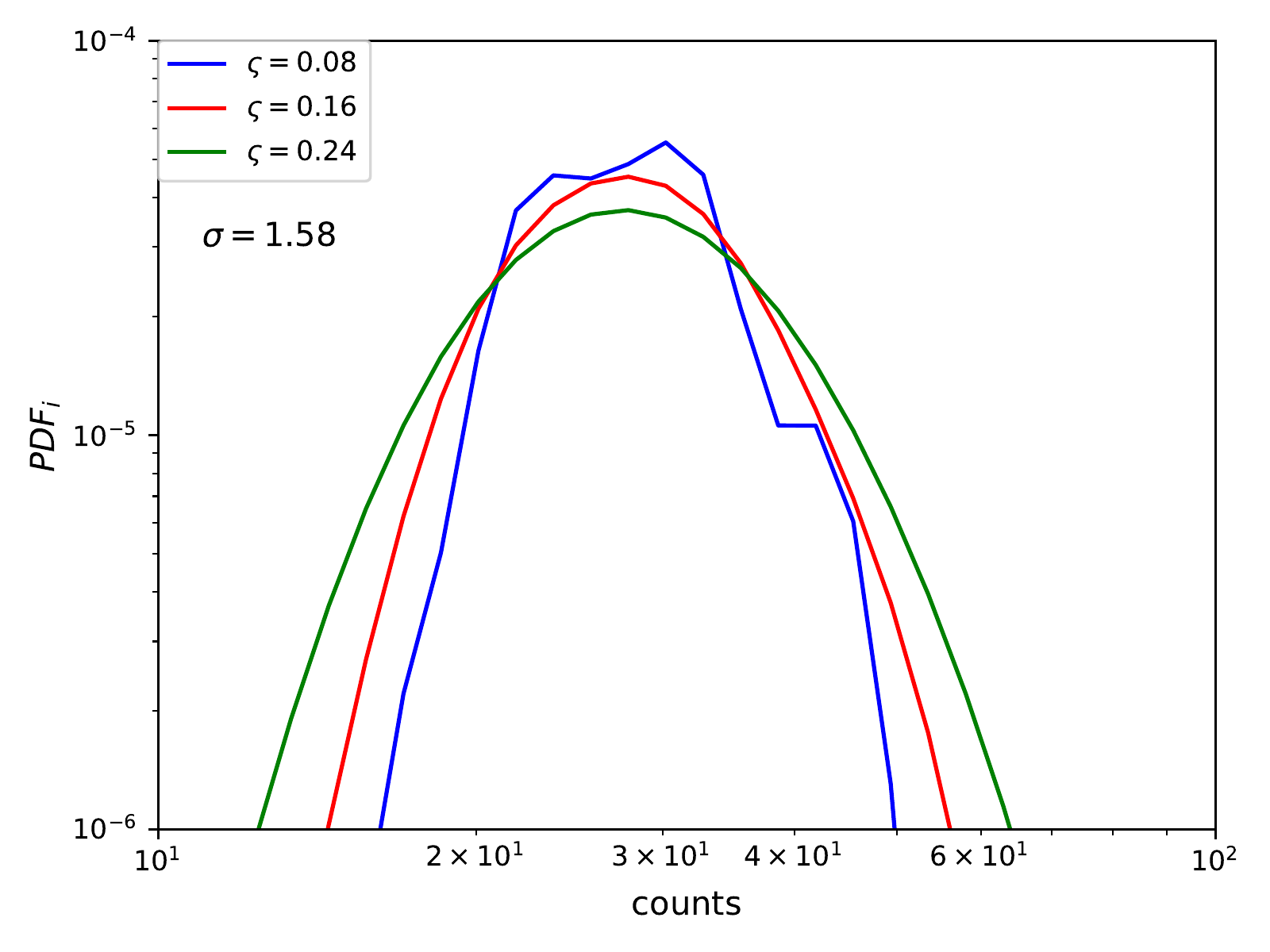}
  \caption{\it Left) PDF evaluated at the point $\vec x = (322.4^\circ, 63.7^\circ)$ having measured count $b=29$  as a function of the smoothing parameter $\sigma$, for 3 different values of $\varsigma$. Right) PDF of the same point vs the counts random variable, while fixing $\sigma=\sigma^*$ for the same 3 values of $\varsigma$.}
\label{fig:PDFi}
\end{figure} 
After the optimisation, the model for the background distribution as a function of the position is completely fixed. 
As we anticipated, we will consider as variable the logarithm of the background counts.
Therefore, accounting for the change of variable, the background PDF writes as: 
\begin{equation}
{\cal B}(\vec x, \ln b)=b\left.{\hat {\cal F}}(\vec x, b)\right|_{\sigma\to \sigma^*,\,\varsigma\to \varsigma^* }\,, 
\label{pdf_bkg}
\end{equation}
where ${\hat{\cal F}}$ is obtained from eq.~(\ref{kernelY}). 
By evaluating this function towards the direction of the dSph $\vec x_d$, we infer the distribution of the background counts there, ${\cal B}(\vec x_d, \ln b)$,
which for ease of notation, in what follows we will denote as ${\cal B}({\ln b}_{d})$. 
For instance, the estimated average number of events simply writes as eq.~(\ref{estim}) evaluated in $\vec x_d$, for $\sigma=\sigma^*$.

\subsection{Energy-dependent analysis}
\label{sec:Ebin}
As described below in sec.~\ref{sec:likelihood}, we perform an energy binned likelihood analysis. The original dataset is distributed into 24 energy bins, from 500 MeV to 500 GeV. However, most of the bins at high energies actually contain zero counts. In particular, for some of the dSphs considered in this analysis, energy bins above $\sim 5$ GeV have none or very little number of counts. Our choice to work with logarithms of the counts forbids us to consider unpopulated bins. For that reason, we perform the analysis by using $N_{\rm bins}=6$, whose upper edges (in units of GeV) are:
\be 
E_{\rm bins} = (0.67, 0.89, 1.19, 1.58, 2.81, 500)~,
\label{Ebins}
\ee
the first 4 energy bins being identical to the ones in the original dataset, which allows one to more directly compare with previous \Fermi-LAT analyses at low DM masses. But from $\sim 3$ GeV upwards, we group all data in one single bin. Its width, which may at first sight appear surprising, is in fact due to the rapidly dropping energy spectrum of the data. This choice certainly has an impact on the derived limits on DM parameters (see the next-to-last paragraph in sec.~\ref{sec:results}), since for sufficiently large DM masses, the limits will be dominated by the last bin above. For those masses, there would be essentially no difference in the limits coming from an energy-integrated analysis, with respect to the choice of eq.~(\ref{Ebins}). We have checked that this is the case for masses $m_{\rm DM}\gtrsim 70$ GeV, following the simplest possible strategy (see {\it case 1} in sec.~\ref{sec:likelihood}). 
 Given the mostly methodological goal of our article, we preferred to stick to the choice  of eq.~(\ref{Ebins}), since it is the one compatible with deriving a desirable estimator of the background (PDF) over the whole sky, i.e. similar {\it relative} errors everywhere, even at the expenses of losing some diagnostic power for the specific case of high-mass DM. An absolute estimator minimising the difference in absolute counts would have penalised the quality of the model in regions with fewer counts and would have allowed one to extract sharper (albeit more fragile) bounds at high-masses. 

\begin{figure}[!htb]
  \centering
  \includegraphics[scale=0.6]{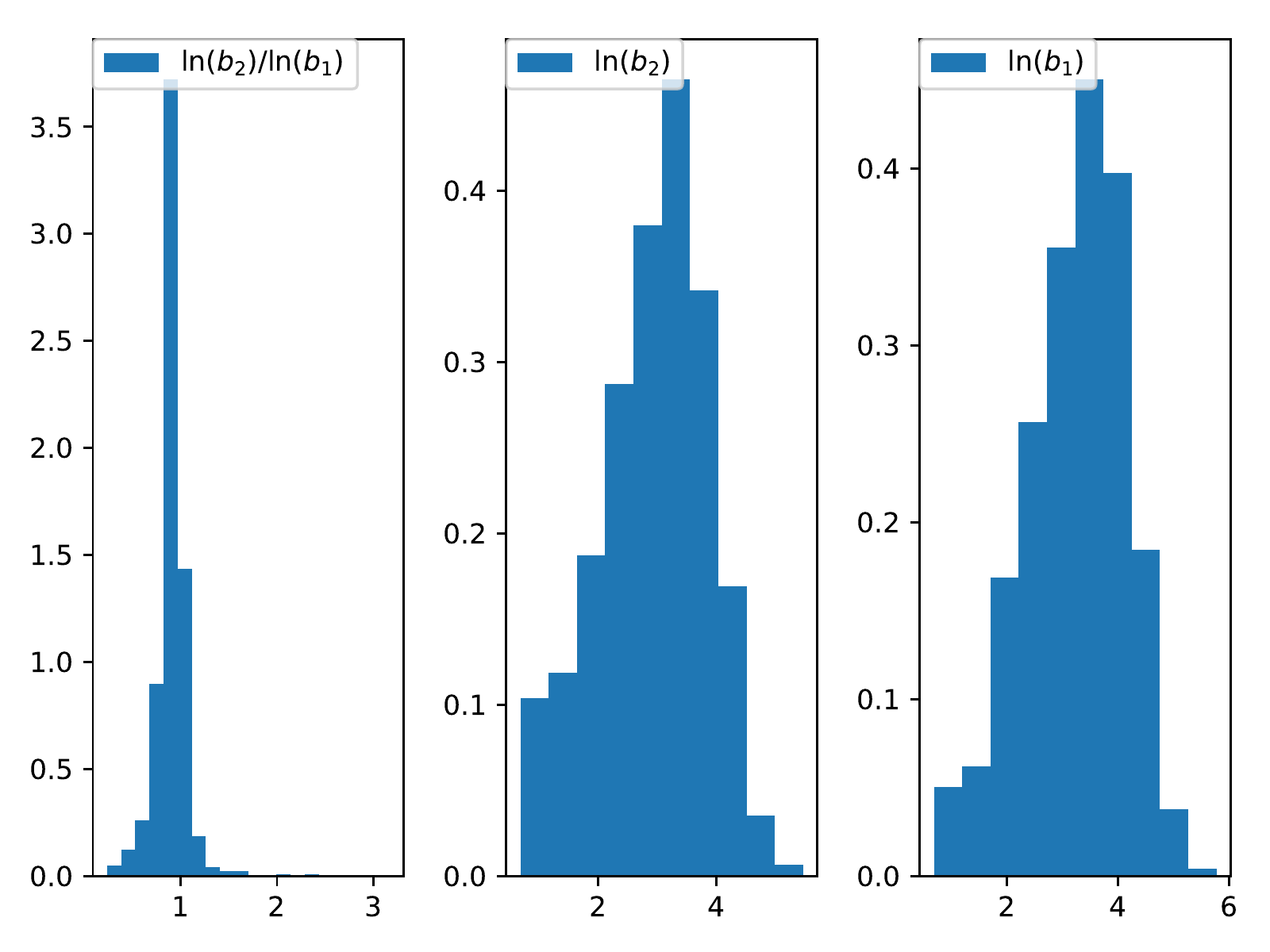} 
  \caption{\it Left) Distribution of the ratio of logarithm of the counts in the second to the first energy bin. Centre) Distribution of the logarithm of the counts in the second energy  bin, and 
  Right) Distribution of the logarithm of the counts in the first energy bin. }
\label{fig:ratios}
\end{figure}

When extending to multiple energy bins, we can define the energy-dependent background PDF as ${\cal B}({\ln b}_{d, e})$.
We work under the hypothesis of perfect correlation of logarithms of counts between energy bins in order to reduce the number
of nuisance parameters and simplify the challenging computational task of profiling over them (see below).
Our reference independent variable is $\ln b_{d,1}$, the logarithm of the background count in the first bin. The optimisation procedure
of $\sigma$ and $\varsigma$ is performed on the statistics of the first bin, and thus the PDF of the background counts in eq.~(\ref{pdf_bkg}) is, strictly speaking, 
the PDF of the background counts in the first bin. 
In order to rescale background predictions to the other bins, we fix the remaining background variables $\ln b_{d,i}$, $i = 2, \dots 6$, so that the 
ratio of the logarithm of their counts to the logarithm of the counts in the first bin, $r_{d,j} \equiv \ln b_{d,j}/\ln b_{d,1}$ is constant and
equivalent to the ratio of the PDFs central values as predicted by eq.~(\ref{estim}) in each bin: $r_{d,j}=\widehat{ \ln  {\sf b}_{d,j}}/\widehat{ \ln {\sf b}_{d,1}}$.
A second working hypothesis enters here, namely that the $\sigma^*,\,\varsigma^*$ only mildly depend on energy (we checked that variations among the first three bins are within $\sim$10\%), and the same kernel of eq.~(\ref{Kest}) can be used in each bin separately.
In order to justify the hypothesis of perfect correlation between bins, we show in fig.~\ref{fig:ratios} the distribution (over the entire dataset of the control regions) of the ratio $\ln b_2/\ln b_1$, 
as compared to the distributions of $\ln b_2$ and $\ln b_1$. As can be observed, the distribution of $\ln b_2/\ln b_1$ is much narrower than the
separate distributions of $\ln b_2$ and $\ln b_1$, which legitimates our hypothesis. We have checked that the same behaviour is obtained for the ratios of other bins. 
Let us stress that none of the above hypotheses are crucial to the method we are applying. They are essentially done to speed-up the computation considerably, but one or several of them may be lifted in an improved treatment.

\section{Likelihood analysis and limits on dark matter}
\label{sec:likelihood}
In order to extract bounds on the DM contribution and also to ease the comparison with \Fermi-LAT results, we follow a frequentist approach detailed below.
\newline\newline\noindent
The most generic form of the likelihood $\mathcal{L}$ at the dSph $d$ in the energy bin $e$ that we will consider is
\be 
{\cal L}_{d,e} (\lambda_{d,e}, \log_{10} J_d,\ln b_{d,e}) = \df{\lambda_{d,e}^{c_{d,e}} e^{-\lambda_{d,e}}}{c_{d,e}!} 
{\cal N}(\log_{10} J_d) {\cal B}(\ln b_{d,e})~,
\label{eq:likelihood}
\ee 
where $c_{d,e}$ is the measured number of counts and $\lambda_{d,e}$ are the expected model counts including signal and background predictions (see section~\ref{sec:DMsignal}):
\be 
\lambda_{d,e}=\lambda_{d,e}(\langle\sigma v\rangle,m_{\rm DM},\log_{10} J_d,\ln b_{d,e}) = 10^{\log_{10} J_d} \, \sigmav \, f_{d,e}(m_{\rm DM}) +e^{\ln b_{d,e}} \,.
\ee
Note that the expressions of $J$'s and $b$'s as exponentials of the logs are purely formal, but make explicit our choice of the (decimal) logarithm of 
the $J$-factor $J_d$ and the (natural) logarithm of background counts $b_{d,e}$ for each dSph 
 as independent variables. Both will be considered in the following as nuisance parameters. The likelihood therefore includes ${\cal N}(\log_{10} J_d)$, a Gaussian distribution for $\log_{10} J_d$:
\be 
{\cal N}(\log_{10} J_d) = \df{1}{\sqrt{2\pi} \sigma^J_d} \exp\left[-\left(\frac{\log_{10} J_d - \overline{\log_{10} J_d}}{\sqrt{2}\sigma^J_d}\right)^2\right] \, , 
\label{eq:likeJ}
\ee
where the $\overline{\log_{10} J_d}$ are the $J$-factors as estimated through spectroscopic measurements, see section~\ref{sec:DMsignal} and tab.~\ref{tab:pred}. 
Analogously, for the background term we use the background PDF ${\cal B}(\ln b_{d,e})$, eq.~(\ref{pdf_bkg}), and its extension to multiple energy bins 
described above.
\newline\newline\noindent
To extract confidence intervals on $\sigmav$, and inspired by the \Fermi-LAT analysis,  we follow the method of profile likelihood as described in ref.~\cite{Ackermann:2013yva}, which we briefly summarise here. As stated by the Wilks theorem \cite{Wilks}, if the data are distributed according to the likelihood function ${\cal L}(\theta_1,...,\theta_N)$, then the test statistics (TS) defined as the maximum log likelihood ratio~\footnote{The Neyman-Pearson lemma states that the log likelihood quantity gives the test statistics with maximum statistical power. }:
\be 
\lambda(\theta_1,\theta_2,...,\theta_k) = -2\ln \df{ {\cal L}(\theta_1,...,\theta_k,\theta^*_{k+1},...,\theta^*_N)}
{{\cal L}^*}\label{WilksTS}~,
\ee
follows, in the large-$N$ limit, a chi-squared distribution with $k$ degrees of freedom. Also note that for the Wilks theorem to apply, the maximum likelihood estimators of the model parameters, $\theta^*_i$, should be normally distributed, which is not guaranteed in general, but was checked by the \Fermi-LAT collaboration in~\cite{Ackermann:2015zua} to be a meaningful approximation for our current application. In the eq.~(\ref{WilksTS}), $\theta_1,...,\theta_k$ represent the parameters of interest, whereas the $\theta_{k+1},...,\theta_N$ represent the nuisance parameters, to profile over, i.e.:
\be 
(\theta^*_{k+1},...,\theta^*_N) = \underset{\theta_{k+1},...,\theta_N}{\rm argmax}\{{\cal L}(\theta_1,...,\theta_k,\theta_{k+1},...,\theta_N) \}
\ee
and 
\be 
{\cal L}^* = {\rm sup}\{{\cal L}(\theta_1,...,\theta_N)\} = {\cal L}(\theta^*_1,...,\theta^*_N)~.
\ee
\noindent
For a fixed DM mass, our parameter of interest is $\sigmav$, and our nuisance variables $(\log_{10} J, \ln b)$. As such, the likelihood eq.~(\ref{eq:likelihood}) becomes:
\bea 
{\cal L}_{d,e} (\sigmav,\log_{10} J_d, \ln b_{d,e}) &=& {\cal P}(\sigmav,\log_{10} J_d, \ln b_{d,e})\, {\cal N}(\log_{10} J_d)\, {\cal B}(\ln b_{d,e})\,.
\label{eq:likelihood2}
\eea
Here ${\cal P}$ refers to the Poisson distribution of the expected counts $\lambda_{d,e}$. For the cases we consider below, the TS will then be specifically:
\be 
\lambda(\sigmav) = -2\ln\df{{\cal L}(\sigmav,\vec \theta(\sigmav))}{{\cal L}^*}~,
\label{eq:TS}
\ee
where the nuisance variables are represented symbolically by $\vec \theta$ and are function(s) of $\sigmav$. We consider the following five cases:
\begin{itemize}
\item {\it Case 1}. For the sake of pedagogical presentation, we start by  fixing both the $J$-factors and the background variables to their estimated central values. For the former, we consider the central value of the corresponding distribution $\overline{\log_{10} J_d}$. 
For the latter, instead, we have shown in sec.~\ref{sec:Method} how to estimate the first moment of the background PDF; we therefore use as central value $\widehat{\ln {\sf b}_{d,e}}$ given by eq.~(\ref{estim}), 
using the same $\sigma^*$ for each bin separately. 
Hence, eq.~(\ref{eq:likelihood2}) reduces to the Poisson term only, such that the total likelihood of dwarf $d$ is:
\be 
{\cal L}_d (\sigmav) = \prod_e^{N_{\rm bins}} {\cal P}(\sigmav,\overline{\log_{10} J_d}, \widehat{\ln {\sf b}_{d,e}}) \, ,
 \label{eq:like_case1}
\ee
where, as discussed in the previous section, we consider $N_{\rm bins}=6$, and extract a limit on $\sigmav$ for each dSph, according to eq.~(\ref{eq:TS}) with no nuisance parameters.

\item {\it Case 2}. We fix the background variable as above, while profiling over the $J$-factors. In this way, the total likelihood at the dSph $d$ becomes:
\be 
{\cal L}_d (\sigmav,\log_{10} J_d) = {\cal N}(\log_{10} J_d)\prod_e^{N_e} {\cal P}(\sigmav, \log_{10} J_d, \widehat{\ln {\sf b}_{d,e}}) \, .
 \label{eq:like_case2}
\ee
The corresponding TS is given by eq.~(\ref{eq:TS}), with $\theta= \log_{10} J_d$.

\item {\it Case 3}. With respect to case 2, we now stack $N_d$ dSphs. The total likelihood is:
\be 
{\cal L}(\sigmav,\boldsymbol{\log_{10} J}) = \prod_d^{N_d}{\cal L}_d (\sigmav,\log_{10} J_d)
\ee
where $\boldsymbol{\log_{10} J}=\{ \log_{10} J_1,...,\log_{10} J_{N_d} \}$, and ${\cal L}_d$ defined in eq.~(\ref{eq:like_case2}).
The nuisance parameters of the corresponding TS of eq.~(\ref{eq:TS}) are $\vec\theta=\boldsymbol{\log_{10} J}$, i.e.
now the profiling is done simultaneously for all $\log_{10} J_d$.
This is our version of the procedure followed by \Fermi-LAT, but using our own background estimate.

\item {\it Case 4}. For a single dSph, we profile {\it simultaneously} over the $J$-factor and the background distribution. 
In our extension to multiple energy bins described in sec.~\ref{sec:Method}, we assumed and justified the (approximate) perfect
correlation among energy bins and we can thus consider only one 
independent background variable, $\ln b_{d,1}$, the other being fixed by the ratios of their central values with respect to the first bin.
We write the total likelihood for the single dSph as:
\bea
{\cal L}_{d} (\sigmav,\log_{10} J_d, \ln b_{d, 1}) &=& {\cal N}(\log_{10} J_d) {\cal B}(\ln b_{d,1})\prod_e^{N_e} {\cal P}(\sigmav,\log_{10} J_d, \ln b_{d,e})~,
\label{eq:like_case4}
\eea
for which the nuisance parameters entering in eq.~(\ref{eq:TS}) of the TS are $\vec\theta=\{\log_{10} J_d,\ln b_{d,1}\}$.

\item {\it Case 5}. Finally, we stack together $N_d$ dSphs, profiling simultaneously on the \Jf-factor and background nuisance parameters.
In this case the total likelihood writes as:
\bea
{\cal L}(\sigmav,\boldsymbol{\log_{10} J}, \boldsymbol{\ln b_1}) = \prod_d^{N_d}{\cal L}_d (\sigmav,\log_{10} J_d, \ln b_{d,1}) \, , 
\label{eq:like_case5}
\eea
where $\boldsymbol{\ln b_1}=\{\ln b_{1,1},..,\ln b_{N_d,1}\}$, and ${\cal L}_d$ is defined in eq.~(\ref{eq:like_case4}).
The nuisance parameters are in this case $\vec\theta=\{\boldsymbol{\log_{10} J},\boldsymbol{\ln b_1}\}$.

\end{itemize}

\section{Results}
\label{sec:results}
In tab.~\ref{tab:pred} we summarise the set of dSphs used in this analysis, together with the adopted \Jf-factors (see also sec.~\ref{sec:DMsignal}), the measured total counts $c$ (integrated in energy, with their Poisson error $\sqrt{c}$) and total expected background counts $\tilde{b}=\exp(\widehat{\ln {\sf b}})$ at the dSph position estimated  according to the procedure detailed in section~\ref{sec:Method}. 

We note how in general the background estimates match well the measured counts within their corresponding errors (``1-2 $\sigma$'s'', bearing in mind the non-Gaussianity of the distributions!) as expected in the case of no DM signal. 
Hence, we compute the limits on $\sigmav$ for the different cases listed in sec.~\ref{sec:likelihood}
and we show below to what extent the determination of background uncertainties relaxes the bounds on the DM signal.

\begin{table}
\begin{center}
\begin{tabular}{| c | c | c | c | c | c |}
\hline
dwarf & name & $\log_{10} J \pm \sigma^J$ & $c \pm \sqrt{c} $ & 
$\tilde{b}$ & $\widehat{\ln\sf b} \pm \Delta(\ln{\sf b})$ \\
\hline\hline
1& Bo\"otes I & $18.2 \pm 0.4$ & 70 $\pm$ 8 & 71.99 & 4.28 $\pm$ 0.2 \\
\hline
2 & Canes Venatici I & $17.4 \pm 0.3$ & 22 $\pm$ 5 & 29.29 & 3.38 $\pm$ 
0.25\\
\hline
3 & Canes Venatici II & $17.6 \pm 0.4$ & 19 $\pm$ 4 & 20.14 & 3.0 $\pm$ 
0.31\\
\hline
4 & Carina & $17.9 \pm 0.1$ & 248 $\pm$ 16 & 259.13 & 5.56 $\pm$ 
0.16 \\
\hline
5 & Coma Berenices$^{*}$ & $19.0 \pm 0.4$ & 27 $\pm$ 5 & 27.61 & 3.32 
$\pm$ 0.51 \\
\hline
6 & Draco$^{*}$ & $18.8 \pm 0.1$ & 221 $\pm$ 15 & 292.8 & 5.68 $\pm$ 0.17\\
\hline
7 & Fornax & $17.8 \pm 0.1$ & 76 $\pm$ 9 & 70.14 & 4.25 $\pm$ 0.2 \\
\hline
8 & Hercules & $16.9 \pm 0.7$ & 348 $\pm$ 19 & 308.32 & 5.73 $\pm$ 0.17\\
\hline
9 & Horologium I & 18.64$\pm$ 0.95  & 125 $\pm$ 11 & 105.21 & 4.66 $\pm$ 
0.21\\
\hline
10 & Hydra II & 16.56 $\pm$ 1.85 & 255 $\pm$ 16 & 275.46 & 5.62 $\pm$ 0.18\\
\hline
11 & Leo I & $17.8 \pm 0.2$ & 164 $\pm$ 13 & 152.69 & 5.03 $\pm$ 0.18\\
\hline
12 & Leo II$^{*}$ &$18.0 \pm 0.2$ & 49  $\pm$  7 &  62.67  &  4.14  
$\pm$  0.21\\
\hline
13 & Leo IV & $16.3 \pm 1.4$ & 118  $\pm$  11  & 121.44  &  4.8  $\pm$  
0.2\\
\hline
14  & Leo V & $16.4 \pm 0.9$ & 129  $\pm$  11  & 112.39  &  4.72  $\pm$  
0.22\\
\hline
15 & Pisces II & 17.9 $\pm$ 1.14 & 178  $\pm$  13 &  174.88  &  5.16  
$\pm$  0.18\\
\hline
16 & Reticulum II & $18.9 \pm 0.6$ & 124  $\pm$ 11  &  120.12  &  
4.79  $\pm$  0.21\\
\hline
17  & Sculptor$^{*}$ & $18.5 \pm 0.1$ & 14  $\pm$  4 &  23.18  &  3.14  
$\pm$  0.34\\
\hline
18 & Segue I$^{*}$ & $19.4 \pm 0.3$ &  158  $\pm$  13 &  138.5  &  4.93  
$\pm$  0.19\\
\hline
19 & Sextans & $17.5 \pm 0.2$ & 179  $\pm$  13  & 166.76  &  5.12  
$\pm$  0.2\\
\hline
20& Tucana II & 18.7 $\pm$ 0.9 & 122 $\pm$  11  & 109.55  &  4.7  $\pm$  
0.18\\
\hline
21  & Ursa Major I &$17.9 \pm 0.5$ & 123  $\pm$  11 &  104.78  &  4.65  
$\pm$  0.19\\
\hline
22 & Ursa Major II &$19.4 \pm 0.4$ & 314 $\pm$  18 &  231.32  &  5.44  
$\pm$  0.17\\
\hline
23 & Ursa Minor$^{*}$ &$18.9 \pm 0.2$ & 187  $\pm$ 14  &  172.49  &  
5.15  $\pm$  0.18\\
\hline
24 & Willman 1 & 19.29 $\pm$ 0.91 & 96  $\pm$  10 &  93.87  &  4.54  
$\pm$  0.18\\
\hline
25 & Grus I & 17.96 $\pm$ 1.93 & 108  $\pm$  10 &  80.37  &  4.39  
$\pm$  0.19\\
\hline\hline
\end{tabular}
\caption{The 25 dSphs to be used in the analysis, with measured 
\Jf-factor (and uncertainties, both in $\log_{10}$ scale) in the 3rd 
column (see sec.~\ref{sec:DMsignal}), the measured total counts with 
their Poisson error (4th column), estimated total background counts (5th 
column), the estimated central values $\ln {\sf b}$ at the dSph position 
with their theoretical errors (last column). Asterisks mark the targets 
we use for the combined limits.}
\label{tab:pred}
\end{center}
\end{table}

\begin{figure}[h]
 \centering
  \includegraphics[scale=0.45]{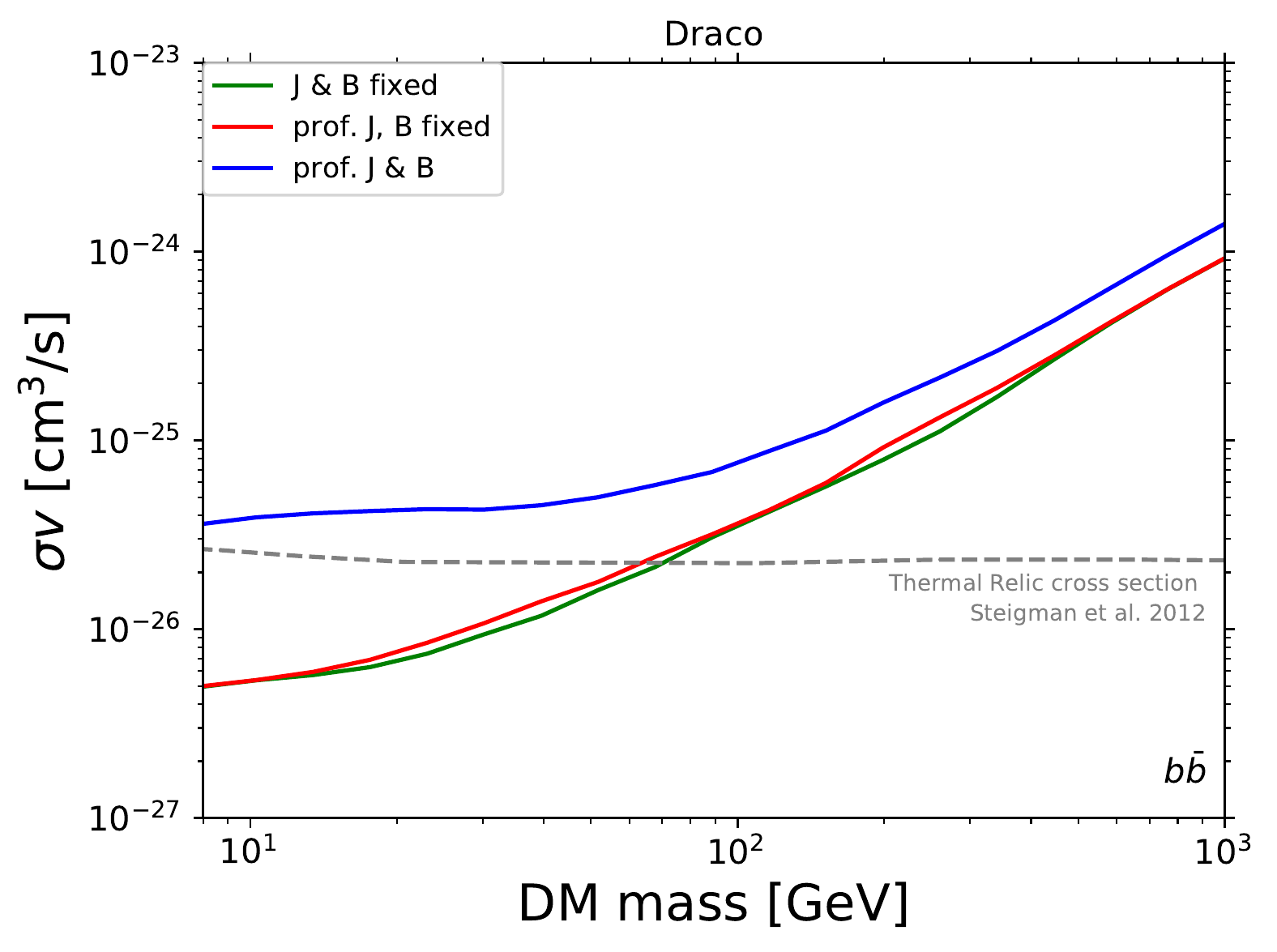}
  \includegraphics[scale=0.45]{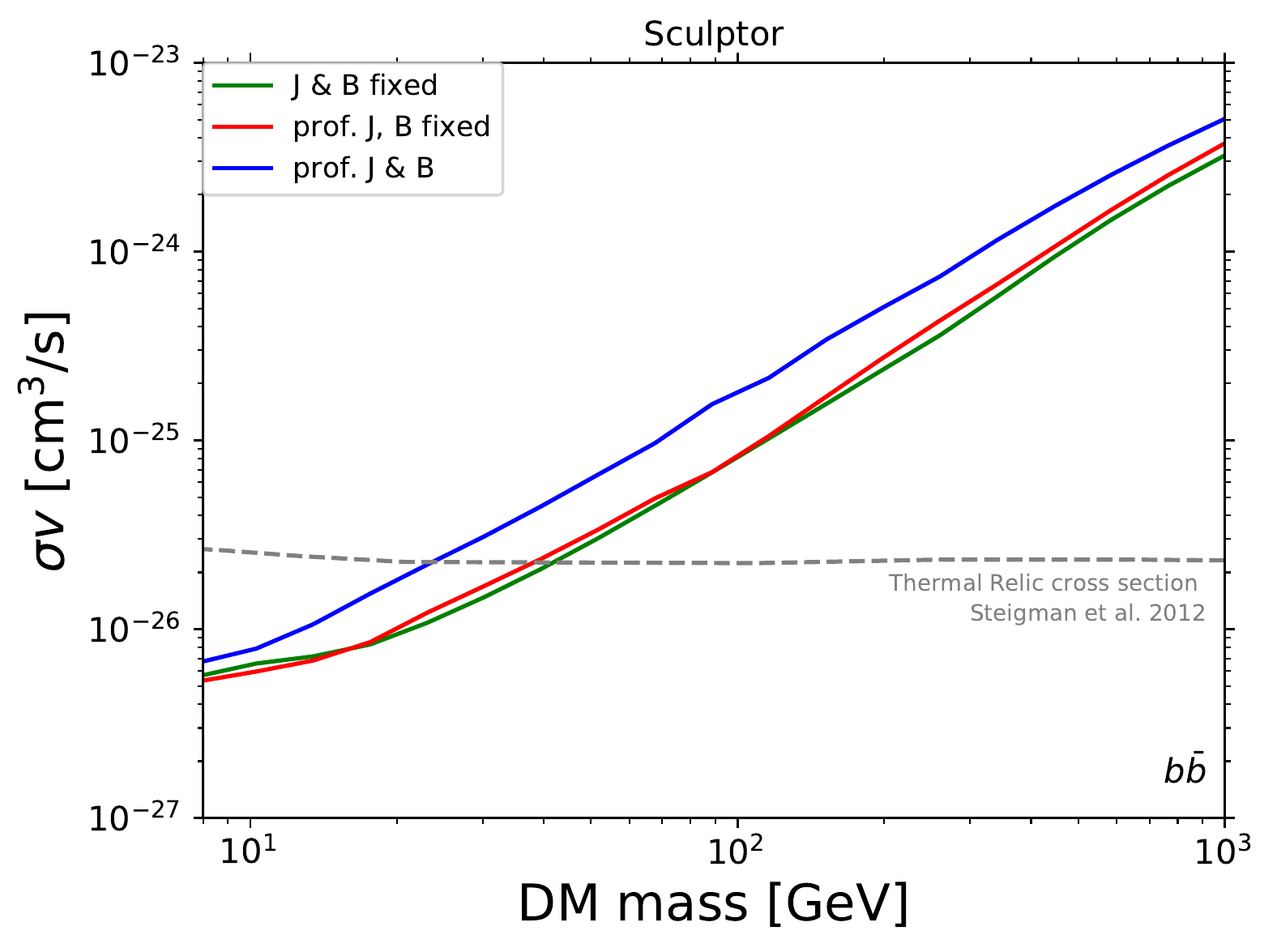}
  \caption{\it Limits on the DM parameters from Draco (left) and Sculptor (right), for cases 1 (green), 2 (red) and 4 (blue).  }
\label{fig:sv_limits}
\end{figure}

\emph{Case 1} does not account for any uncertainty on the \Jf-factor and background nuisance parameters and,
as such, will lead to the strongest constraints on $\sigmav$ (see fig.~\ref{fig:sv_limits} for an illustration of the behaviour of two of the most important dwarfs).
We note that even in this very simplistic case, the ranking of dSphs based of their constraining power
does not strictly reflect the ranking based on their \Jf-factor only -- as sometimes assumed for the selection of
best targets for pointed instruments data taking. 
As an example, the \Jf-factor of Segue I is higher than the \Jf-factor of Coma Berenices, but the fact that the background underestimates the counts at the former whereas for Coma Berenices it coincides with the observed value causes the bound from the latter to be stronger.    We find that, at low DM masses, the four most constraining dSphs according to \emph{case 1} are: Willman 1, Coma Berenices, Segue I and Draco.
This ranking is determined by a subtle interplay between background estimate and \Jf-factor central 
value, which both enter in the determination of the Poisson term of the likelihood. At high masses the ranking changes and Draco provides the stronger bound instead. This is because for the last energy bins (sensitive to the largest DM masses) the expect backgrounds of Draco continue to overestimate the counts, whereas the rest, even if having larger \Jf-factors, has a background underestimating the counts\footnote{Note that the bin-per-bin counts are not shown in tab. \ref{tab:pred}.}.

Profiling over nuisance parameters adds additional freedom in the likelihood and the optimal values for 
\Jf-factors and background counts might, in principle, be different from their central values.
To understand the impact of profiling over the selected nuisance parameters, in fig.~\ref{fig:sv_limits}, 
we show the limits for Draco (left panel) and Sculptor (right panel).
When profiling over the \Jf-factor uncertainties only, \emph{case 2}, the limits weaken with respect to \emph{case 1} as expected, with the degradation depending on the width of 
the $\log_{10} J$ distribution. If, additionally, we profile also over background uncertainties, \emph{case 4}, in the case of the single dSph the limits further
degrades. However, it is important to note that this degradation is not the same for all dSphs and that the ranking of the most constraining 
dSphs can be in principle altered by this procedure.
This is indeed what we see happening when
comparing Draco with, for example, Sculptor. In the case of profiling over \Jf-factor only,  
 Draco still provides stronger limits than  Sculptor because the variation of the \Jf-factor of the latter is not enough to drive the limits further down. On the other hand, when profiling also over background, at low masses  Draco becomes less constraining. The reason is than in the first bin the deficit is less significative than for Sculptor, and the background distribution makes possible to accommodate for larger DM contributions.
At high DM masses, instead, Draco always dominates the constraints for the same reason as in \emph{case 1}.

\begin{figure}[h]
  \centering
  \includegraphics[scale=0.45]{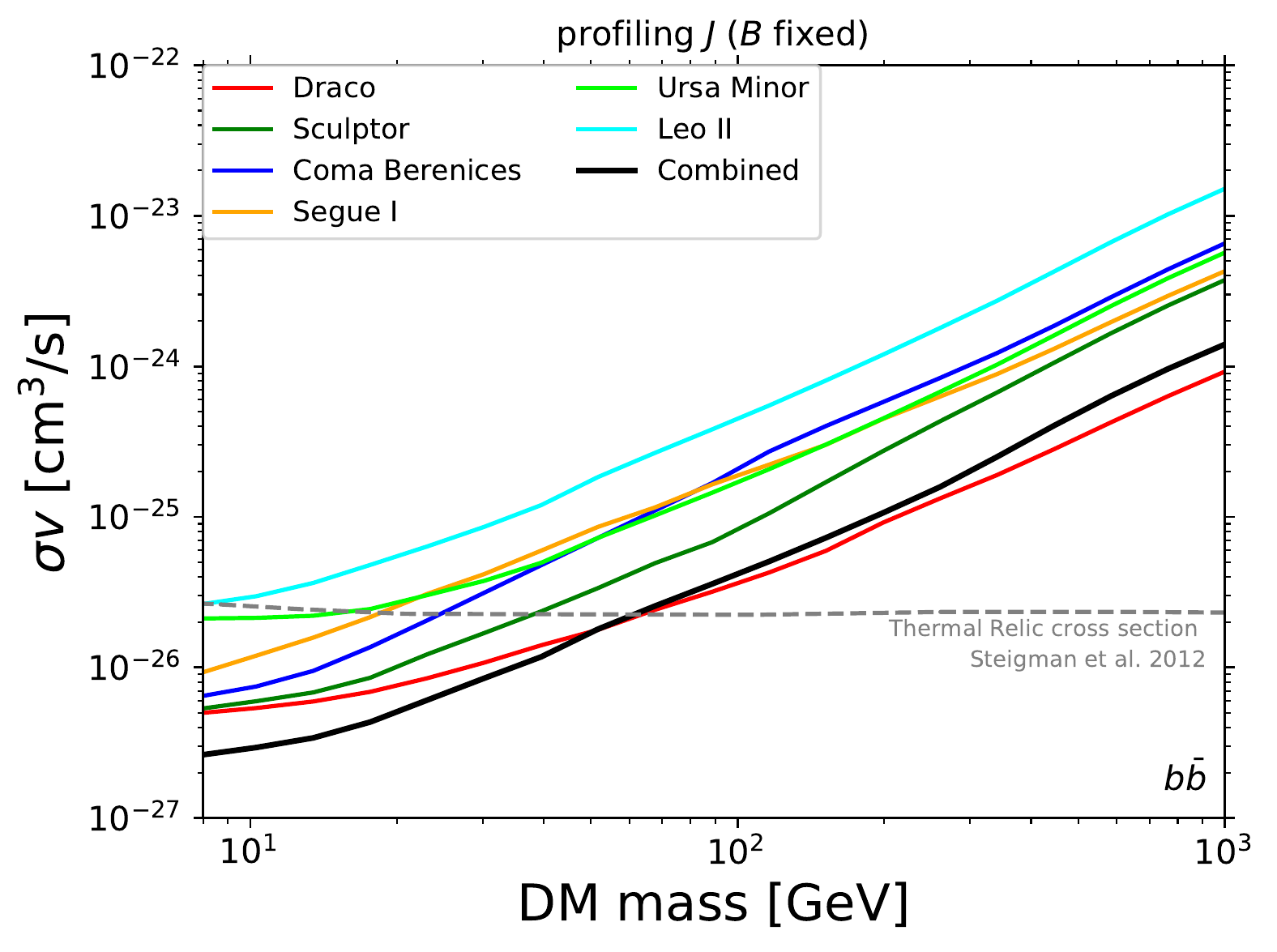}
  \includegraphics[scale=0.45]{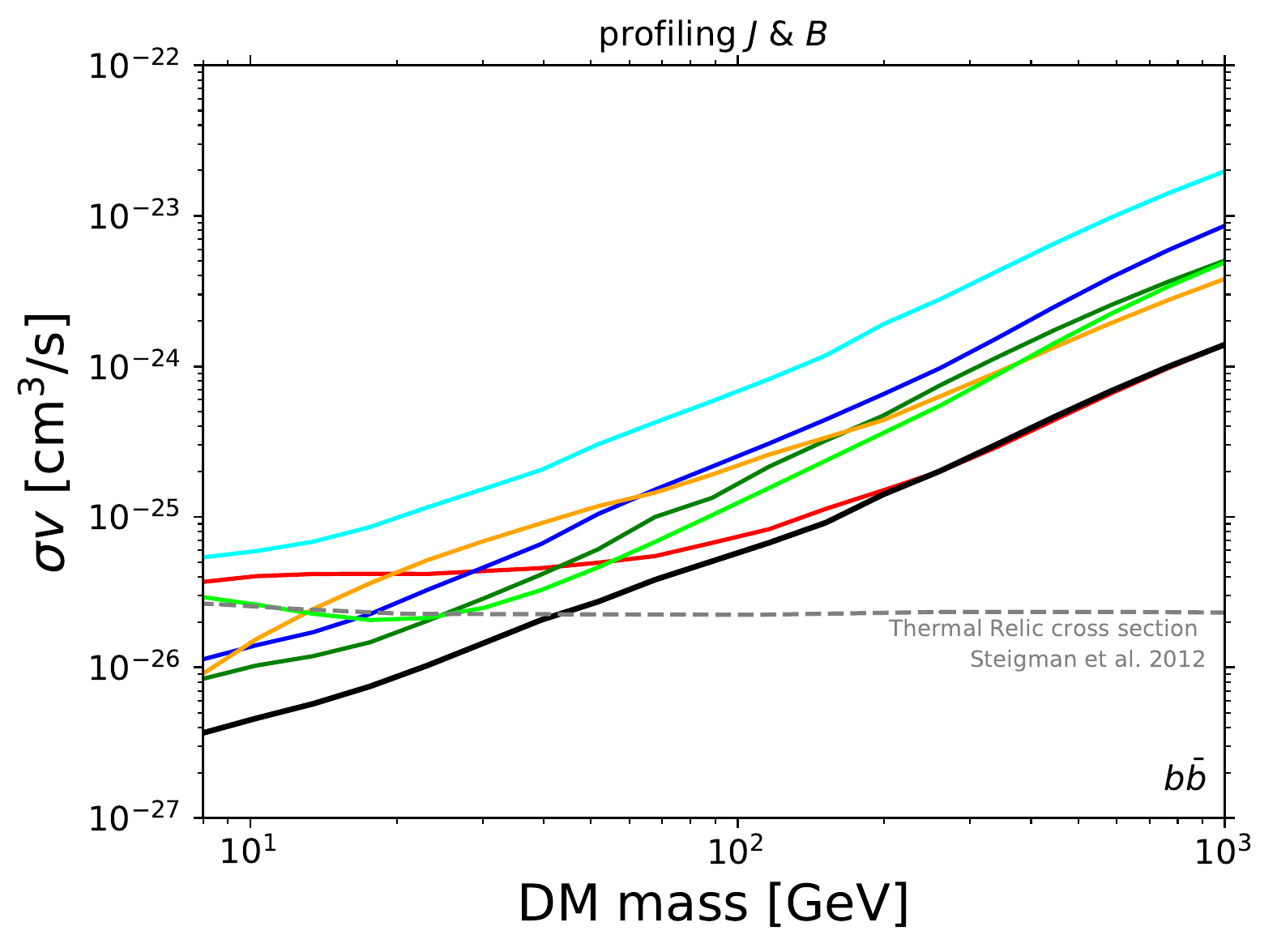}
  \caption{\it Left) Limits on the DM parameters coming from the 6 strongest dSphs of case 2 (coloured lines), and corresponding stacked limit (case 3, black line). 
  Right) Analogous to left panel, but profiling also over background uncertainties (case 4 and 5).}
\label{fig:sv_limits1}
\end{figure}

In fig.~\ref{fig:sv_limits1}, left panel, we display the results separately for the six most constraining dSphs\footnote{We note that 
the selected 6 objects are sufficiently far in position to be considered as independent measurements since our background determination is anyhow local, as explained above.} once we profile 
over the \Jf-factor nuisance parameter, with the background fixed to its central value. 
As explained before, the freedom introduced by profiling over \Jf-factor uncertainties not only degrades the limits with respect to \emph{case 1},
but also changes the ranking of the most constraining dSphs. In this case, Coma Berenices gives a weaker limit than Sculptor
-- while for \emph{case 1} it was the opposite. This can be understood by looking at the uncertainty on $\log_{10} J$, which is larger for Coma Berenices, enabling more freedom in adjusting the \Jf-factor preferred value for this dSph during the profiling procedure.
The results of the stacking procedure with profiling over \Jf-factor, \emph{case 3},
is depicted by the black solid line:Draco dominates by far the combined constraint at all masses, 
additionally, at low masses, its interplay 
with Sculptor and Coma Berenices becomes important.

\begin{figure}[h]
 \centering
  \includegraphics[scale=0.7]{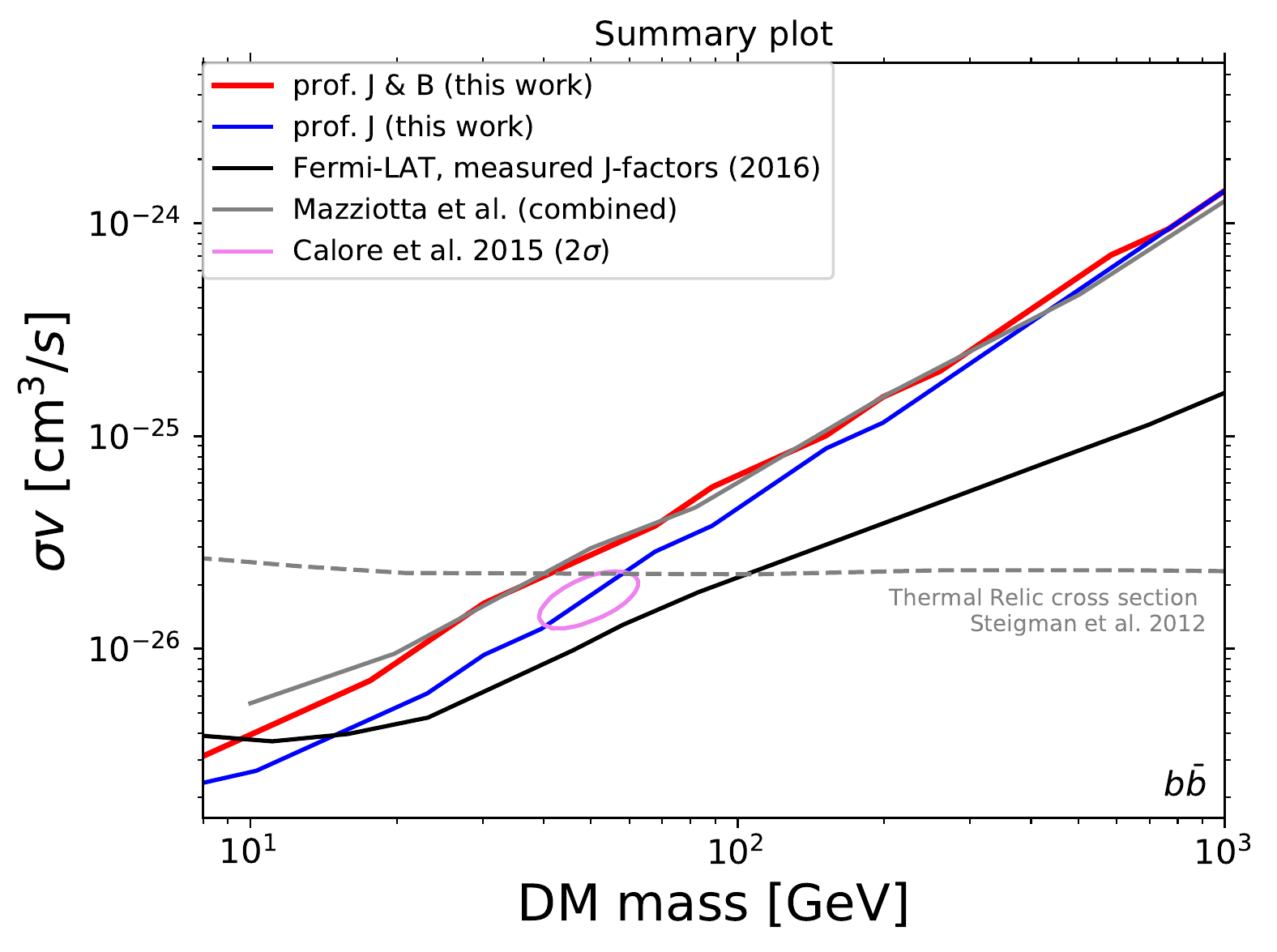}
  \caption{\it Comparison of current limits on the DM parameters: the main result of this work when profiling over 
  \Jf-factor (blue line) and also background uncertainties (red line), \Fermi-LAT limits using a sample of 16 dSphs with measured \Jf-factor from ref.~\cite{Fermi-LAT:2016uux}
  (black line), and results from a more data-driven approach in ref.~\cite{Mazziotta:2012ux} (grey line).
  The purple contour represents the 2$\sigma$ best-fit region for the DM interpretation of the Galactic Centre Excess in ref.~\cite{Calore:2014xka}.
  }
\label{fig:sv_limits_compare}
\end{figure}

In fig.~\ref{fig:sv_limits1}, right panel, we show the effect of profiling also over background uncertainties (\emph{case 4}), on single dSphs for the same six objects as in left panel, as well as their combined limit (\emph{case 5}). In general, the limits degrades further
and the ranking of the most constraining dSphs is not the same as in the case of profiling over \Jf-factor only, especially at low masses, 
 similarly to what explained above in the comparison between Draco and Sculptor.
In this case, the stacked limit is a factor $\sim 1.5$ weaker than the one for \emph{case 3}, and this difference is dominated by the variation of the limits from Draco. We stress that the {\it proper inclusion of background uncertainties has a significant impact on the DM constraints, and, as such, should be fully included in analyses of dSphs.}
\newline\newline\noindent
Finally, we compare our results with the ones in the literature in figure~\ref{fig:sv_limits_compare}. 
In order to make a comparison as fair as possible, rather than comparing with the baseline results reported in~\cite{Fermi-LAT:2016uux}, we  
compare our results with the \Fermi-LAT stacked limit obtained using a sample of 16 dSphs with measured \Jf-factor (red curve in figure 10 of ref.~\cite{Fermi-LAT:2016uux}), i.e. with an analysis more similar in spirit to the previous \Fermi-LAT publication on the subject~\cite{Ackermann:2015zua}. 
Our combined limit when profiling only over \Jf-factor uncertainties (blue curve) is comparable with the latest \Fermi-LAT results (black curve) at  masses around 15 GeV, while getting better at low masses and performing less and less well at high masses. The improvement at small masses was anticipated (see section~\ref{introd}), since this regime is the most background-limited, and an optimised background estimate may produce some improvement. Sub-leading effect responsible for the difference are the slightly (by one year) enlarged statistics of the data set and the different set of dSphs used compared with the combined limit presented in~\cite{Fermi-LAT:2016uux}. Overall, this indicates that our background determination method performs equally well or better than the \Fermi-LAT pipeline, while being more general and flexible.
At high masses, \Fermi-LAT limits are stronger, again as expected for the reasons discussed in detail in section~\ref{sec:Ebin}: This is notably related to the binning of high-energy data and is ultimately reflecting the limited statistics rather than the background modeling. Note that our limits turn out to be stronger than  or at worst comparable to the ones coming from more data-driven approaches, e.g. ref.~\cite{Mazziotta:2012ux}. 
This is reassuring, at very least in view of the optimisation that we performed, as well as the slightly improved statistics.

Another lesson learned is that when profiling also over the background (red line in fig.~\ref{fig:sv_limits_compare}), the constraints degrade substantially, around a factor 1.5 to 2,
with respect to the case of profiling only over the \Jf-factor uncertainties (blue line in the same figure), which shows the importance of including the background uncertainties in the analysis.  
We emphasise that this difference is not anecdotal: for instance, at face value, the blue curve in fig.~\ref{fig:sv_limits_compare} might appear in tension with DM interpretations of the so-called Galactic Centre Excess, see e.g.~\cite{Calore:2014xka}. We are of course aware that this might be a too strong conclusion, given the different methodologies and systematics involved in the two results: the tension may be reduced or vanish altogether once accounting for example for uncertainties on the DM distribution in the Milky Way~\cite{Calore:2014nla,Benito:2016kyp}. Nonetheless, one sees by eye that the red line is in no tension at all proving that, alone, a more sound treatment of the background uncertainty in the dSph might be already crucial
to relax away the tension.

\section{Discussion and conclusions}
\label{sec:conclusions}
We have presented a general data-driven method to estimate the distribution of the gamma-ray background at any sky position, consistently in the whole sky, relying on \Fermi-LAT measurements. 
The method developed here is based on a non-parametric, unbiased estimate of the underlying background PDF~\cite{parzen1962}, which may depends, in principle, on all counts in the regions of the sky identified as signal-free.
The optimisation of the PDF parameters through an MLE method has shown that the estimate of the background at any point is dominated by its closest neighbours, an outcome indicating a rather noisy input (photon counts in the sky). 

Equipped with the full PDF for the background, as a major application, we performed an energy binned likelihood analysis of the gamma rays coming from a sample of 25 dSphs and derived the 95\% C.L.~upper limits on the DM annihilation cross-section, by {\it simultaneously profiling over  background and \Jf-factors uncertainties}. 

In order to make contact with previous results in the literature, we considered first the case of profiling over \Jf-factor only while fixing the background to its expectation. In this case, at masses
$m_{\rm DM} \sim 15$ GeV our limits are comparable with the ones obtained by the \Fermi-LAT collaboration using a similar data set and dSphs sample,
while improving over them at lower masses, and worsening at masses $m_{\rm DM}\gg 15$ GeV. Both effects are understood: The low-$m_{\rm DM}$ improvement originates mostly from the optimized background estimate method, while the high-$m_{\rm DM}$ worsening is related to the binning of high-energy data (made necessary by our specific choice, eq.~\ref{kernelY}) and is ultimately reflecting the limited statistics, rather than the background estimate method {\it per se}.

On the other hand, by taking into account also the background uncertainties we observe a noticeable degradation of the limits with respect to the case where only the uncertainties on the \Jf-factor are considered. The limits are weaker by a factor 1.5 to 2 depending on the DM mass, such that a thermal  relic DM is excluded for DM masses up to $m_{\rm DM}\sim 40$ GeV, compared to the $m_{\rm DM}\sim 60$ GeV exclusion of the latter case (or $\sim$ 100 GeV in the case of the \Fermi-LAT limits), see figure~\ref{fig:sv_limits_compare}. Although our study is closer in spirit to the analyses performed in refs.~\cite{GeringerSameth:2011iw,Mazziotta:2012ux,Boddy:2018qur}, in the sense of not relying on astrophysical modelling, in addition it accounts for both background uncertainties and \Jf-factor uncertainties in the determination of the limits. Also, it removes some arbitrariness in the choice of the control region to assess the background, since in our case this is consistently determined via an optimisation procedure. In particular, we compare our results with those from \cite{Mazziotta:2012ux}, and obtain fully comparable limits when accounting for the uncertainties in the background predictions, while definitively stronger bounds if limiting ourselves to account for \Jf-factor uncertainties.

An additional advantage of our method is that it applies to any position in the sky: equipped with the current model, the estimate of the background counts at the position of a newly discovered dSph can be done immediately, depending only upon the position of the new dSph. The procedure is in that sense much simpler that what required in the \Fermi-LAT pipeline, where the definition of a new control window and background estimate would be necessary.
Making sensitivity predictions to DM detection and limits setting for arbitrary spatial dSphs distributions would for instance be much more straightforward in our method.

There are of course several ways to improve and generalise the methodology proposed here. We list here a few examples: 
\begin{itemize} 
\item One can introduce extra parameters of the background PDF over which to perform the MLE. An obvious extension would consist for instance of a different $\sigma$ for the distance in latitude and longitude.
\item One can switch to a (more) local (as opposed to global) optimisation of the PDF, meaning that one can minimise directly $-2\ln {\hat {\cal F}}$ at each grid point (or, in an intermediate phase, in coarse grained patches of the sky). 
This will lead to position-dependent optimal values $\sigma^*$, $\varsigma^*$, notably at the training sample points $\vec x_i$. 
These meta-parameters $\sigma_i^*$, $\varsigma_i^*$ can be interpolated into the signal region properly accounting for spatially-dependent background PDF.
\item In section~\ref{sec:Ebin} we described some simplification to the treatment of multiple energy bins. Some or all of these approximations can be lifted, since they are not linked to fundamental limitations of the method, and the bounds may be further refined. This also holds for additional effects, such as the dependence of the results upon the ``pixel size'', which we kept fixed, but could be subject to further optimisation. 
\end{itemize} 

The main difficulty that we found in attaining a highly satisfactory description of the background, and a common problem to many procedures, is that the background is not only a smooth function of the position, but contains unresolved sources responsible for localised ``pixel excesses'', themselves distributed in the sky in an unknown way. Their coarse-grained distribution could be smoothly varying with Galactic coordinates (as for instance expected for millisecond pulsars) or be largely isotropic (as for extragalactic BL Lac objects), and each sub-population has in general different energy spectra. We note that one way to use the model to improve in that respect is to try to identify these populations by studying ``excesses'' with respect to the expectations in the non-signal regions. This may also suggest further refinements and generalisations in the kernel or functions used to describe the PDF, in order to model also a non-Poissonian contribution to the counts and identify possible sub-threshold point-source candidates.

Regardless of these refinements, there is a clear message emerging from the results presented here: \emph{a proper data-driven inclusion of background uncertainties in the derivation of DM upper limits is not only possible, but very much needed for a fair comparison with limits from other targets}, where theoretical background uncertainties are traditionally more explored than in dSphs analyses. It is worth noting that the possible tension between hints for a Galactic Centre Excess, if interpreted as due to DM, and dSph bounds, as reported e.g.~in ref.~\cite{Ackermann:2015zua}, crucially depends also on the treatment of the astrophysical background uncertainties in the dSphs, as we have explicitly shown. 

\section*{Acknowledgements}
We would like to warmly acknowledge E. Fernandez-Martinez and M. Blennow, as well as F. Costanza for the enlightening discussions about the statistical procedure. We also thank P. Aubert for useful discussions on computer code optimisation.  We finally acknowledge  A.~Drlica-Wagner for  discussion about the adopted \Fermi-LAT procedure. F.C.~and B.Z.~would like to thank all participants of the 
``Accelerating the Search for Dark Matter with Machine Learning'' workshop (15 -- 19 January 2018, Leiden, NL) for useful and inspiring discussions about machine learning methods and their application to dark matter challenges. B.Z. is supported by the ``Investissements d' avenir'' program of the French ANR, Labex ``ENIGMASS''.

\clearpage
\bibliographystyle{JHEP}
\bibliography{biblio}

\end{document}